\documentclass[10pt, conference, compsocconf]{IEEEtran}
\IEEEoverridecommandlockouts

\usepackage{moresize}
\usepackage{amsmath,amssymb,mathrsfs}

\usepackage[inline]{enumitem}
\usepackage[normalem]{ulem}
\usepackage{graphicx}
\usepackage{caption,subcaption}

\newcommand{\bft}{\textsc{Bft}}
\newcommand{\pbft}{PBFT}

\newcommand{\zyzzyva}{Zyzzyva}

\newcommand{\expodb}{{ResilientDB}}

\newcommand{\pbc}{\textsc{Pbc}}

\newcommand{\MName}[1]{\textsc{#1}}
\newcommand{\GETS}{:=}
\newtheorem{example}{Example}

\usepackage{tikz}

\begin{document}

\title{Permissioned Blockchain Through the Looking Glass: Architectural and Implementation \\Lessons Learned}

\author{\IEEEauthorblockN{Suyash Gupta$^{*}$\thanks{$^*$Both authors have equally contributed to this work.} \hspace{4mm}Sajjad Rahnama$^{*}$ \hspace{4mm}Mohammad Sadoghi}
\IEEEauthorblockA{
MokaBlox LLC. \\
Exploratory Systems Lab \\
University of California, Davis\\
}}

\maketitle

\begin{abstract}
Since the inception of Bitcoin, the distributed systems community has shown interest 
in the design of efficient blockchain systems.
However, initial blockchain applications (like Bitcoin) attain very low throughput, which 
has promoted the design of {\em permissioned} blockchain systems.
These permissioned blockchain systems 
employ classical Byzantine-Fault Tolerant (\bft{}) protocols to reach consensus. 
However, existing permissioned blockchain systems still attain low throughputs 
(of the order $10$K txns/s).
As a result, existing works blame this low throughput on the associated \bft{} protocol 
and expend resources in developing optimized protocols.
We believe such blames only depict a one-sided story. 
In specific, we raise a simple question: {\em can a well-crafted system based on a 
classical \bft{} protocol outperform a modern protocol?}
We show that designing such a well-crafted system is possible and illustrate 
that even if such a system employs a three-phase protocol, it can outperform another system 
utilizing a single-phase protocol.
This endeavor requires us to dissect a permissioned blockchain system and 
highlight different factors that affect its performance.
Based on our insights, we present the design of our enterprise-grade, 
high-throughput yielding permissioned blockchain 
system, \expodb{}, that employs multi-threaded deep pipelines, to balance tasks at a replica, 
and provides guidelines for future designs.

\end{abstract}

\section{Introduction}
Since the inception of {\em blockchain}~\cite{bc-processing,blockbench}, the distributed systems 
community has renewed its interest in the age-old design of 
Byzantine-Fault Tolerant (\bft{}) systems. 
At the core of any blockchain applications is a \bft{} algorithm that 
ensures all the replicas of this blockchain 
application reach a {\em consensus}, that is, 
agree on the order for a given client request, even if some of 
the replicas are byzantine~\cite{pbft,zyzzyva,disc-mbft,byz-agreement,hyperledger-fabric}.

Surprisingly, even after a decade of its introduction and publication of several prominent research works, 
the major use-case of blockchain technology remains as a crypto-currency.
This leads us to a key observation: {\em Why have
blockchain (or \bft{}) applications seen such a slow adoption?} 

The low throughput and high latency are the key reasons why \bft{} 
algorithms are often ignored.
Prior works~\cite{easyc,quecc,qstore,easyc-extend} have shown that the traditional 
distributed systems can achieve throughputs of the order $100$K transactions
per second
while the initial blockchain applications, such as
Bitcoin~\cite{bitcoin} and Ethereum~\cite{ether}, have throughputs
of at most ten transactions per second.
Such low throughputs do not affect the users of these applications, as 
the aim of these applications is to promote an alternative currency, which 
is unregulated by any large corporation, 
that is, anyone can join, and the identities of the participants are kept hidden 
({\em open membership}).
Evidently, this open-membership property has also led to several attacks on these 
applications~\cite{blockbench,bitcoin-hack-1,sybill}.

This led to industry-grade {\em permissioned} blockchain systems, 
where only a select group of users, some of which may be untrusted, can participate~\cite{hyperledger-fabric}.
However, the throughputs of current permissioned blockchain applications are
still of the order $10$K transactions per second~\cite{hyperledger-fabric,caper,ahl}.
Several prior works blame the low throughput and scalability of a permissioned blockchain system 
on to its underlying \bft{} consensus algorithm~\cite{blockbench,zyzzyva,hotstuff,ahl}. 
Although these claims are not false, we believe they only represent a one-sided story.

\begin{figure}[t]
   \centering
   \includegraphics[width=0.6\columnwidth]{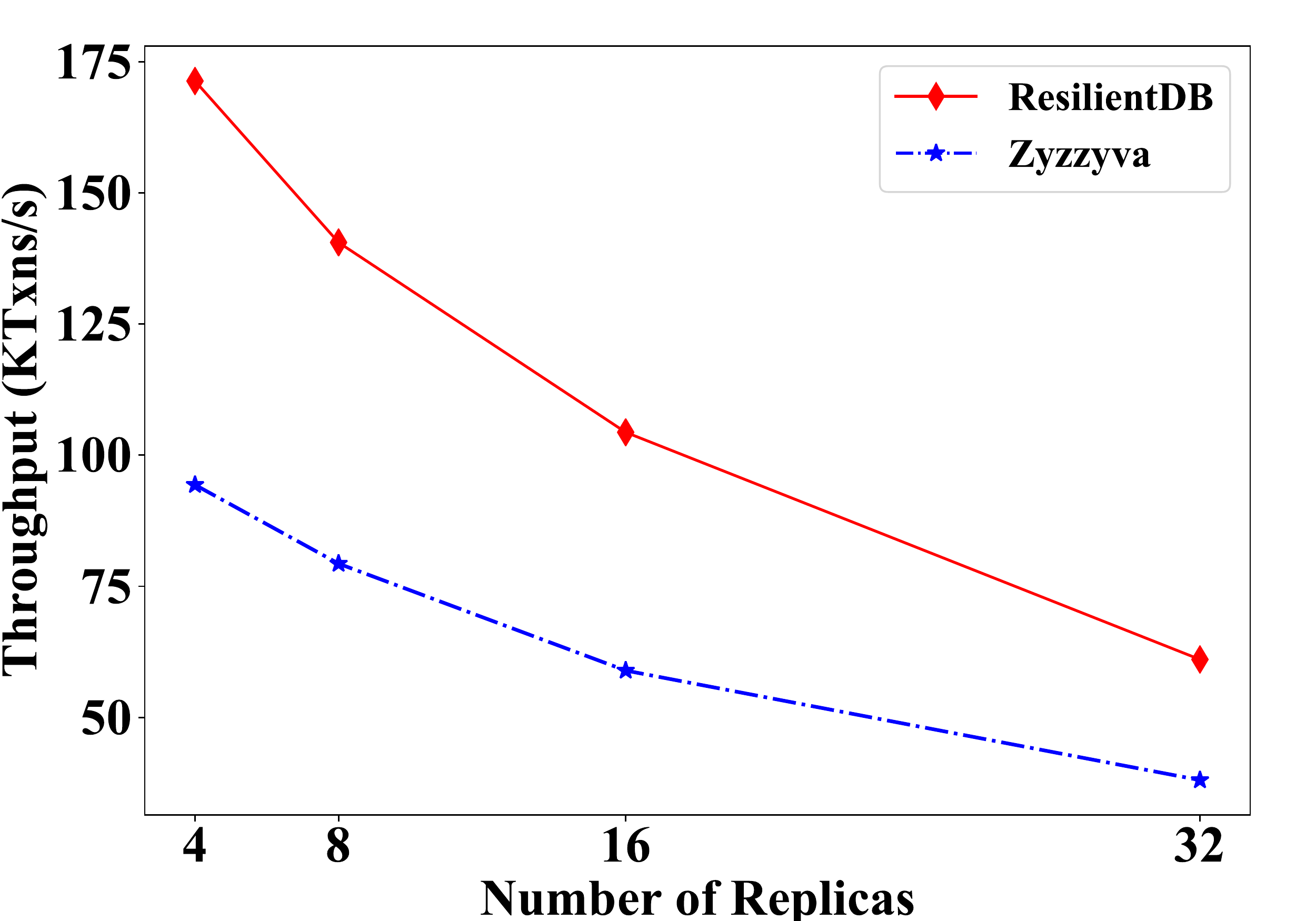}
   \caption{Two permissioned applications 
	employing distinct \bft{} consensus protocols 
	($80K$ clients used for each experiment).}
   \label{fig:compare-tput}
\end{figure}

We claim that the low throughput of a blockchain system 
is due to missed opportunities during its design and implementation.
Hence, we want to raise a question: 
{\em can a well-crafted system-centric architecture based on a classical \bft{} protocol 
outperform a protocol-centric architecture?}
Essentially, we wish to show that even a slow-perceived classical \bft{} protocol, such as \pbft{}~\cite{pbft}, 
if implemented on skillfully-optimized blockchain fabric, 
can outperform a fast niche-case and optimized for fault-free consensus, \bft{} protocol, such as \zyzzyva{}~\cite{zyzzyva}.  
We use Figure~\ref{fig:compare-tput} to illustrate such a possibility.
In this figure, we measure the throughput of an optimally designed 
permissioned blockchain system (\expodb{}) and intentionally make it employ the slow \pbft{} protocol. 
Next, we compare the throughput of \expodb{} against a {\em protocol-centric} permissioned blockchain system 
that adopts practices suggested in BFTSmart~\cite{bftsmart} and  
employs the fast \zyzzyva{} protocol.
We observe that the {\em system-centric} design of \expodb{}, even after employing 
the three-phase \pbft{} protocol (two of the three phases require quadratic communication among the replicas) 
outperforms the system having a single-phase linear protocol \zyzzyva{}.
Further, \expodb{} achieves a throughput of $175$K 
transactions per second, scales up to $32$ replicas, and attains up to $79\%$ 
more throughput.

This paper is aimed at illustrating that the design and architecture of the 
underlying system are as important as optimizing \bft{} consensus.
Decades of academic research and industry experience 
has helped the community in designing efficient distributed applications~\cite{future-cloud,architecture-nextgen,big-data-plat,tbook}.
We use these principles to illustrate the design of a 
high-throughput yielding permissioned blockchain fabric, \expodb{}.
In specific, we {\em dissect} existing permissioned blockchain systems,
identify different performance bottlenecks, and illustrate mechanisms to 
eliminate these bottlenecks from the design. 
For example, we show that even for a blockchain system, ordering of transactions can 
be easily relaxed without affecting the security.
Further, most of the tasks associated with transaction ordering 
can be extensively parallelized and pipelined.
A highlight of our other observations:
\begin{itemize}
\item Optimal batching of transactions 
	can help a system gain up to $66\times$ throughput.
\item Clever use of cryptographic signature schemes can increase throughput by $103\times$.
\item Employing in-memory storage with blockchains can yield up to $18\times$ throughput gains.
\item Decoupling execution from the ordering of client transactions can increase throughput by $10\%$.
\item Out-of-order processing of client transactions can help gain $60\%$ more throughput.
\item Protocols optimized for fault-free cases can result in a loss of $39\times$ throughput under failures.
\end{itemize}
These observations allow us to perceive \expodb{} as a reliable test-bed to implement and evaluate  
enterprise-grade blockchain applications.
\footnote{
\expodb{} is available and open-sourced at https://resilientdb.com.}
We now enlist our contributions:

\begin{itemize}
\item We dissect a permissioned blockchain system and enlist different factors that affect its performance.
\item We carefully measure the impact of these factors and present ways to mitigate the effects of 
	these factors.
\item We design a permissioned blockchain system, \expodb{} that yields high throughput, incurs 
low latency, and scales even a slow protocol like \pbft{}.
\expodb{} includes an extensively parallelized and deeply pipelined architecture that efficiently balances the load at a replica.
\item We raise {\em eleven} questions and rigorously evaluate our \expodb{} 
platform in light of these questions.
\end{itemize}

{\bf \em Note on this work:}
This paper is not aimed at designing efficient \bft{} consensus protocols, for which 
there already exists an extensive literature~\cite{pbft,zyzzyva,algorand,rapidchain,blockplane}.
Further, this work does not aim at benchmarking open-membership and permissioned blockchain systems, 
as done by Blockbench~\cite{blockbench}.
Moreover, we do not advocate the use of any specific \bft{} protocol or permissioned blockchain system, but 
instead perform an in-depth analysis of a single permissioned blockchain system, to uncover insights  
that can help both researchers and practitioners to build next-generation blockchain fabrics. 




\section{Background and Related Work}
\label{s:back}
Before laying down the foundation for efficient design, we first analyze existing 
literature and practices 
in the domain of permissioned blockchain.

\subsection{BFT Consensus}

\begin{figure}[t]
	\centering
    	\includegraphics[width=0.5\columnwidth]{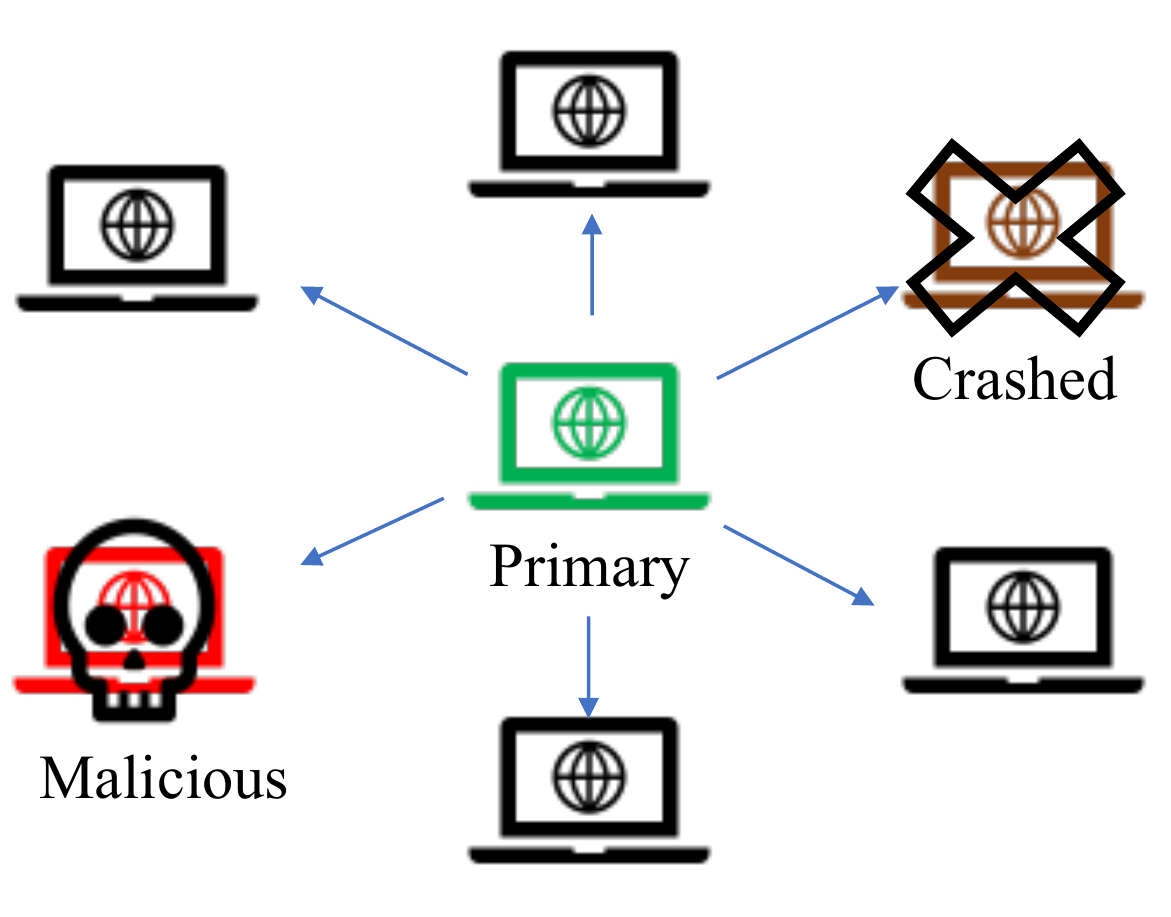}
	\caption{This diagram illustrates a set of replicas of which some may be malicious or have crashed. One replica is designated as the primary, which leads the consensus on the received client request among the backup replicas.}
	\label{fig:consensus}
\end{figure}

At the core of any blockchain application is a \bft{} consensus protocol,
which states that given a client request and a set of replicas, 
some of which could be byzantine, 
the non-faulty replicas would agree on the order for this client request. 
We use Figure~\ref{fig:consensus} to schematically represent consensus.

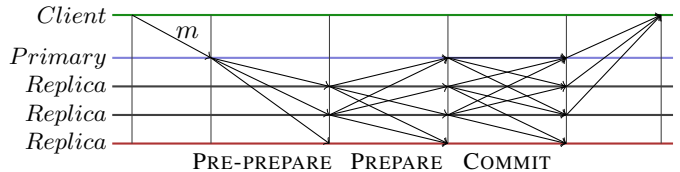
\begin{figure}[t!]
    \centering
    \begin{tikzpicture}[yscale=0.38,xscale=1.05]
        \draw[thick,draw=black!75] (0.75, 4.5) edge[green!50!black!90] ++(7.1, 0)
                                   (0.75,   0) edge[red!60!black!80] ++(7.1, 0)
                                   (0.75,   1) edge ++(7.1, 0)
                                   (0.75,   2) edge ++(7.1, 0)
                                   (0.75,   3) edge[blue!70!black!50] ++(7.1, 0);

        \draw[thin,draw=black!75] (1,   0) edge ++(0, 4.5)
                                  (2, 0) edge ++(0, 4.5)
                                  (3.5,   0) edge ++(0, 4.5)
                                  (5, 0) edge ++(0, 4.5)
                                  (6.5,   0) edge ++(0, 4.5)
                                  (7.7,   0) edge ++(0, 4.5);

        \node[left] at (0.8, 0) {\small $Replica$};
        \node[left] at (0.8, 1) {\small $Replica$};
        \node[left] at (0.8, 2) {\small $Replica$};
        \node[left] at (0.8, 3) {\small $Primary$};
        \node[left] at (0.8, 4.5) {\small $Client$};

        \path[->] (1, 4.5) edge node[above=-4pt,xshift=6pt,label] {$m$} (2, 3)
                  (2, 3) edge (3.5, 2)
                         edge (3.5, 1)
                         edge (3.5, 0)
                           
                  (3.5, 2) edge (5, 0)
                           edge (5, 1)
                           edge (5, 3)

                  (3.5, 1) edge (5, 0)
                           edge (5, 2)
                           edge (5, 3)

		  (5, 3) edge (6.5, 1)
			 edge (6.5, 2)
			 edge (6.5, 3)
                         
                  (5, 2) edge (6.5, 0)
                         edge (6.5, 1)
                         edge (6.5, 3)

                  (5, 1) edge (6.5, 0)
                         edge (6.5, 2)
                         edge (6.5, 3)

		  (6.5, 3) edge (7.7, 4.5)
		  (6.5, 2) edge (7.7, 4.5)
		  (6.5, 1) edge (7.7, 4.5)
                           ;
                           
        \node[below,label] at (2.65, 0) {\small \MName{Pre-prepare}};
        \node[below,label] at (4.35, 0) {\small \MName{Prepare}};
        \node[below,label] at (5.75, 0) {\small \MName{Commit}};
    \end{tikzpicture}%
    \caption{The three-phase \pbft{} protocol.}
    \label{fig:pbft-protocol}
\end{figure}

{\em \bf \pbft{}}~\cite{pbft} is often described as the first \bft{} protocol 
to allow consensus to be incorporated by practical systems.
\pbft{} follows the primary-backup model where one replica is designated as the {\em primary} 
and other replicas act as the backup. 
\pbft{} only guarantees a successful consensus among $n$ replicas 
if at most $f$ of them are byzantine, where $n \geq 3f+1$.

When the primary replica receives a client request, it assigns it a sequence number 
and sends a $\MName{Pre-prepare}$ message to all the backups to execute this request 
in the sequence order (refer to Figure~\ref{fig:pbft-protocol}).
Each backup replica on receiving the $\MName{Pre-prepare}$ message from the primary shows its agreement 
to this order by broadcasting a $\MName{Prepare}$ message.
When a replica receives $\MName{Prepare}$ message from at least $2f$ distinct backup 
replicas, then it achieves a guarantee that a {\em majority} of the non-faulty replicas are aware 
of this request. 
Such a replica marks itself as {\em prepared} and broadcasts a $\MName{Commit}$ message.
Next, when this replica receives $\MName{Commit}$ messages from $2f+1$ distinct replicas, 
then it achieves a guarantee for the order of this request, as a majority of the replicas must have also 
prepared this request. 
Finally, this replica executes the request and sends a response to the client.

{\em \bf More Replicas.} It is evident from the \pbft{} protocol that 
any system relying on \pbft's design for consensus would be expensive.
Hence, several other efficient \bft{} designs (each with its limitation) 
have emerged lately~\cite{tutorial-middleware}.
For instance, Q/U~\cite{qu-bft} attempts to reduce \bft{} consensus to a single-phase 
through the use of $5f+1$ replicas, but cannot handle concurrent requests.
HQ~\cite{hq} builds on top of Q/U and permits concurrency only if the transactions are non-conflicting.

{\em \bf Speculative Execution.} \zyzzyva{}~\cite{zyzzyva} introduces speculative execution 
to the \bft{} protocols to yield a linear \bft{} protocol. 
In \zyzzyva's design, as soon as a backup replica receives a request from the primary, it executes the 
request and sends a response to the client. 
Hence, a replica does not even wait to confirm that the order is the same across all the replicas.
\zyzzyva{} requires just one phase, so its design helps to gauge the maximum throughput that 
can be attained by a \bft{} protocol.
However, if the primary is malicious, \zyzzyva{} needs the help of its good clients to ensure a correct order. 
If the clients are malicious, then \zyzzyva{} is unsafe until a good client participates.
Further, \zyzzyva's fast case requires a client to receive a response 
from all the $3f+1$ replicas before it marks a request complete.
Prior works~\cite{aadvark} have shown that just one failure is enough 
to lead \zyzzyva{} to very low throughput.
A recent protocol, {\em Proof-of-Execution} (PoE) tries to remove the limitations of \zyzzyva{} and outperforms \pbft{}~\cite{poe}. 
However, PoE also requires one phase of quadratic communication among its replicas.



{\bf \em Multiple Primaries:} 
Several protocols~\cite{disc-mbft,hotstuff,multibft-system} suggest dedicating 
multiple replicas as primaries to gain higher throughput. 
Multiple primaries can boost the throughput when the system is not limited by resources and network. 
Further, multiple primary protocols require coordination for the execution of their requests in 
the correct order.

\subsection{Chain Management}
A blockchain is an immutable ledger that consists of a set of {\em blocks}. 
Each block contains necessary information regarding the executed transaction and the 
{\em previous} block in its chain.
The data about the previous block helps any blockchain achieve immutability.
The $i$-{th} block in the chain can be represented as:
$B_i \GETS \{k, d, v, H(B_{i-1}) \}$

This block $B_i$ contains the sequence number ($k$) 
of the client request, the digest ($d$) of the request, 
the identifier of the primary $v$ who initiated the consensus, and 
the hash of the previous block, $H(B_{i-1})$. 
%
In each blockchain application, every replica independently maintains its 
copy of the blockchain.
Prior to the start of consensus, the blockchain of each replica has no element.
Hence, it is initialized with a {\em genesis} block~\cite{bc-processing}.
The genesis block is marked as the first block in the chain and contains dummy data. 
For instance, a genesis block can contain the hash of the identifier of the first primary, $H(\mathbb{P})$.

\begin{figure}[t]
	\centering
    	\includegraphics[width=0.9\columnwidth]{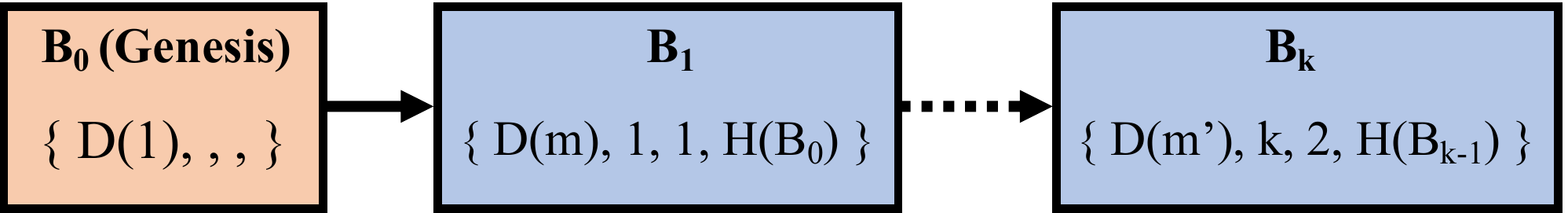}
	\caption{A formal representation of the blockchain.}
	\label{fig:example-blockchain}
\end{figure}

\subsection{Alternative Blockchain Architectures}
To improve the throughput attained by a permissioned blockchain application, researchers 
have also explored alternate architectures and designs to manage the blockchain.


{\bf \em DAG.}
Since the common data-structure in any blockchain application is the ledger, 
several systems incorporated a directed-acyclic graph to record the 
client transactions~\cite{caper,conflux,spectre}.
As a blockchain application expects a single order for all the transactions across
all the replicas, so a DAG-based design allows replicas working on non-conflicting 
transactions to simultaneously record multiple transactions.
However, a DAG-based design would require the merge of branches of a DAG once there 
are conflicting transactions, which in turn necessitates communication between all the replicas.

{\bf \em Sharding.}
Another approach to extract higher throughput from a blockchain system is
to employ sharding~\cite{caper,ahl,parblockchain}.
Sharding splits the records accessed by the clients into several distinct partitions, 
where each partition is maintained by a set of replicas.
Although sharding helps an application to attain high throughput when client transactions 
require access to only one partition, multi-partition transactions are expensive as they 
can require up to two additional phases to ensure safety.

{\bf \em Geo-Scale Clustering.} 
Several protocols have suggested clustering replicas in the vicinity~\cite{steward,geobft}. 
For example, GeoBFT~\cite{geobft} facilitates running the \pbft{} protocol in each cluster in parallel.
Although this design yields high-throughput, it reduces fault-tolerance as each cluster needs to have $3f+1$ replicas.


{\bf \em Other Permissioned Systems.} Several other permissioned blockchain systems
such as Hyperleder Fabric~\cite{hyperledger-fabric}, MultiChain~\cite{multichain}, and 
Tendermint~\cite{tendermint} have proposed high-level system architecture to achieve 
high throughput. 
For instance, Hyperledger Fabric presents a distinct paradigm of executing transactions 
first and then ensuring they have a valid order, while Tendermint advocates reliance on a
synchronous setting to attain higher throughput.
Despite these exciting principles, these works miss the low-level system details, 
which is the main focus of this work.
Further, several existing \bft{} protocols have employed BFTSmart~\cite{bftsmart} 
as a standard implementation for \pbft{}.
This is noteworthy as BFTSmart associates a non-pipelined architecture with \pbft{} 
and avoids other design optimizations for the sake of design simplicity.

\section{Dissecting Permissioned Blockchain}
\label{s:dissect}
Most of the strategies we discussed in the previous section focussed at: 
(i) optimizing the underlying \bft{} consensus algorithm, and/or
(ii) restructuring the way a blockchain is maintained.
We believe there is much more to render in the design of a permissioned blockchain 
system beyond these strategies.
Hence, we identify several other key {\em factors} that reduce the throughput 
and increase the latency of a permisisoned blockchain system or database.

\textbf{Single-threaded Monolithic Design.}
There are ample opportunities available in the design of a permissioned blockchain application 
to extract parallelism.
Several existing permissioned systems provide minimal to no discussion 
on how they can benefit from the underlying hardware or cores~\cite{caper,ahl,rapidchain}.
Due to the sustained reduction in hardware cost 
(as a consequence of Moore's Law~\cite{moore-law}), it is easy for each replica 
to have at least {\em eight} cores.
Hence, by parallelizing the tasks across different threads and pipelining several transactions,
a blockchain application can highly benefit from the available computational power.

\textbf{Successive Phases of Consensus.}
Several works advocate the benefits of performing consensus on one request at a time~\cite{caper,streamchain}, 
while others promote aggregating client requests into large batches~\cite{hyperledger-fabric,bitcoin}.
We believe there is a communication and computation trade-off that needs to be analyzed before
reaching such a decision.
Hence, an optimal batching limit needs to discovered.

\textbf{Decoupling Ordering and Execution.}
On receiving a client request, each replica of a permissioned blockchain application 
has to order and execute that request.
Although these tasks share a dependency, it is a useful design practice to separate 
them at the physical or logical level. 
At the physical level, distinct replicas can be used for execution. 
However, such an approach would incur additional communication costs. 
At the logical level, distinct threads can be asked to process requests in parallel, but 
additional hardware cores would be needed to facilitate such parallelism.
In specific, a single entity performing both ordering and execution loses an opportunity 
to gain from inherent parallelism.

\textbf{Strict Ordering.}
Permissioned blockchain applications rely on \bft{} protocols, which necessitate ordering 
of client requests in accordance with linearizability~\cite{pbft,linearizability}.
Although linearizability helps in guaranteeing a safe state across all the replicas, 
it is an expensive property to achieve. 
Hence, we need an approach that can provide linearizability but is inexpensive. 
We observe that permissioned blockchain applications can benefit from delaying the ordering of 
client requests until execution. 
This delay ensures that although several client requests are processed in parallel, the result of 
their execution is in order.

\textbf{Off-Memory Chain Management.}
Blockchain applications work on a large set of records or data. 
Hence, they require access to databases to store these records. 
There is a clear trade-off when applications store data in-memory 
or on an off-the-shelf database. 
Off-memory storage requires several CPU cycles to fetch data~\cite{comparch-book}.
Hence, employing in-memory storage can ensure faster access, which in turn can lead 
to high system throughput.

\textbf{Expensive Cryptographic Practices.}
Blockchain applications expect the exchange of several messages among the participating replicas and the clients, of which some may be byzantine.
Hence, each blockchain application requires strong cryptographic constructs that allow a client or a replica to 
validate any message.
These cryptographic constructs find a variety of uses in a blockchain application:
(i) To sign a message.
(ii) To verify an incoming message.
(iii) To generate the digest of a client request.
(iv) To hash a record or data.

To sign and verify a message, a blockchain application can employ either symmetric-key cryptography 
or asymmetric-key cryptography~\cite{cryptobook}.
Although symmetric-key signatures, such as Message Authentication Code (MAC), 
are faster to generate than asymmetric-key signatures, such as Digital Signature (DS), 
DSs offer the key property of non-repudiation, which is not guaranteed by MACs~\cite{cryptobook}.
Hence, several works suggest using DSs~\cite{hyperledger-fabric,caper,ahl,rapidchain}.
However, a cleverly designed permissioned blockchain system can skip using DSs for a majority 
of its communication, which in turn will help increase its throughput.
For generating digests or hash, a blockchain application needs to employ 
standard Hash functions, such as SHA256 or SHA3, which are secure. 

\section{\expodb{} Permissioned Blockchain Fabric}
\label{s:impl}

\begin{figure}[t]
	\centering
	\includegraphics[width=0.7\columnwidth]{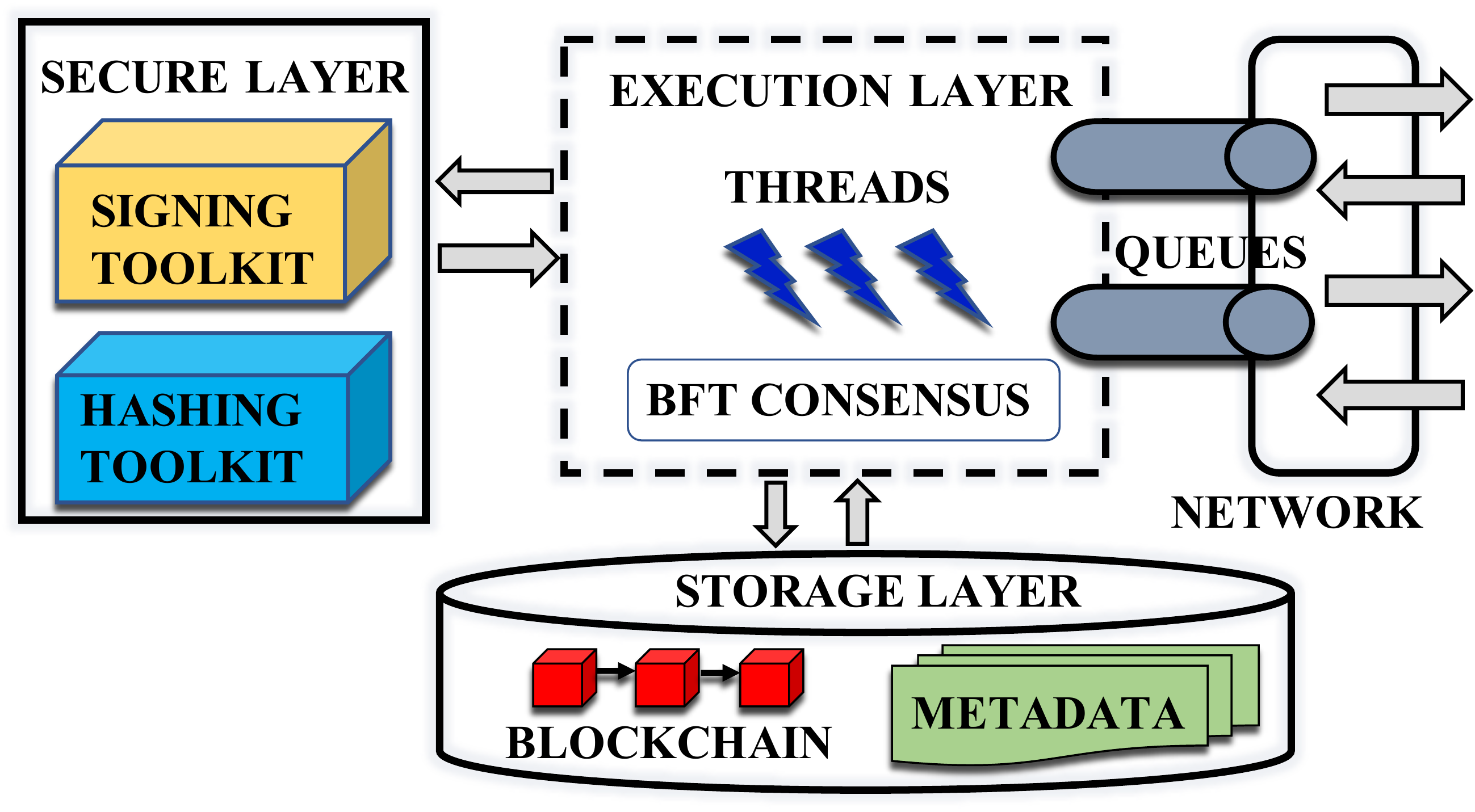}
	\caption{\expodb{} Architecture.}
	\label{fig:expodb}
\end{figure}

We now present our \expodb{} fabric, which incorporates our insights 
and fulfills the promise of an efficient permissioned blockchain system.
%
In Figure~\ref{fig:expodb}, we illustrate the overall architecture of \expodb{}, which
lays down an efficient client-server architecture. 
At the {\em application layer}, we allow multiple clients to co-exist, each of which creates 
its own requests. 
For this purpose, they can either employ an existing benchmark suite or design a 
{\em Smart Contract} suiting to the active application. 
Next, clients and replicas use the {\em transport layer} to exchange messages across the network.
\expodb{} also provides a {\em storage layer} where all the metadata corresponding to a request and 
the blockchain is stored.
At each replica, there is an {\em execution layer} where the underlying consensus protocol is run 
on the client request, and the request is ordered and executed.
During ordering, the {\em secure layer} provides cryptographic support.

Since our aim is to present the design of a high-throughput permissioned blockchain system, 
for the rest of the discussion we use the simple yet robust \pbft{} protocol 
(explained in Section~\ref{s:back}) for reaching 
consensus among the replicas.
Note that succeeding insights also apply to other \bft{} protocols.

\subsection{Multi-Threaded Deep Pipeline}
For implementing \pbft{}, we require \expodb{} to follow the primary-backup model.
On receiving a client request, the primary replica must initiate \pbft{} 
consensus among all the backup replicas and ensure all the replicas 
execute this client request in the same order.
Note that depending on the choice of \bft{} protocol, \expodb{} can be molded to 
adopt a different model (e.g. leaderless architecture).
%
%

\begin{figure}[t]
	\centering
	\includegraphics[width=\columnwidth]{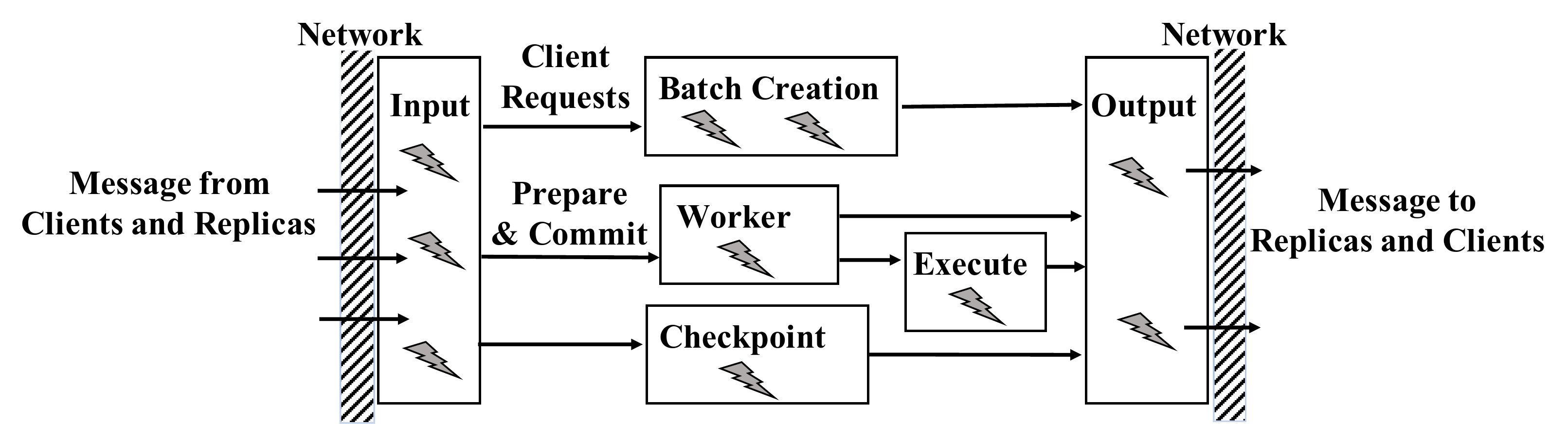}
	\caption{Schematic representation of the multi-threaded deep-pipelines at each \expodb{} replica. 
	The number of threads of each type can be varied depending on the requirements of the underlying consensus protocol.}
	\label{fig:prim-pipe}
\end{figure}

In Figure~\ref{fig:prim-pipe}, we 
illustrate the threaded-pipelined architecture of \expodb{} replicas.
We permit increasing (or decreasing) the number of threads of each type. 
In fact one of the key goals of this paper is to study the effect of 
varying these threads on a permissioned blockchain.
With each replica, we associate multiple {\em input} 
and {\em output} threads.
In specific, we balance the tasks assigned to the input-threads, by 
requiring one input-thread to solely receive client 
requests, while two other input-threads to collect messages sent by other replicas.
\expodb{} also balances the task of transmitting messages between the two output-threads 
by assigning equal clients and replicas to each output-thread.
To facilitate this division, we need to associate a distinct {\em queue} with 
each output-thread.

\subsection{Transaction Batching}
\expodb{} allows both clients and replicas to batch their transactions. 
Using an optimal batching policy can help mask communication and consensus costs.
A client can send a burst of transactions as a single request to the primary. 
Examples of applications where a client may batch multiple transactions are stock-trading, 
monetary-exchanges, and service level-agreements.
The primary replica can also aggregate client requests together to significantly reduce the number of times a consensus 
protocol needs to be run among the replicas.

\subsection{Modeling a Primary Replica}
To facilitate efficient batching of requests, \expodb{} 
associates multiple {\em batch-threads} with the primary replica.
When the primary replica receives a batch of requests 
from the client, it treats it as a single request.
The input-thread at the primary assigns a 
monotonically increasing sequence number to each incoming client request 
and enqueues it into the common queue for the batch-threads.
To prevent contention among the batch-threads, we design the 
common queue as {\em lock-free}.
But {\em why have a common queue?} 
This allows us to ensure that any enqueued request is consumed as 
soon as any batch-thread is available.

Each batch-thread also performs the task of verifying the signature of the client request.
If the verification is successful, then it creates a batch and names it as the
 $\MName{Pre-prepare}$ message. 
\pbft{} also requires the primary to generate the digest of the client request and send 
this digest as part of the $\MName{Pre-prepare}$ message. 
This digest helps in identifying the client request in future communication.
Hence, each batch-thread also hashes a batch and marks this hash as a digest.
%
Finally, the batch-thread signs and enqueues the corresponding $\MName{Pre-prepare}$ message  
into the queue for an output-thread.

Apart from the client requests, the primary replica receives
$\MName{Prepare}$ and $\MName{Commit}$ messages from backup replicas.
As the system is partially asynchronous, so the primary may receive both the $\MName{Prepare}$ 
and $\MName{Commit}$ messages from a backup replica $X$ before the $\MName{Prepare}$ 
message from a backup $Y$.
{\em How is this possible?} The replica $X$ could have received sufficient number of $\MName{Prepare}$ 
messages (that is $2f$) before the primary receives $\MName{Prepare}$ from replica $Y$ (total number of replicas are $n = 3f+1$).
In such a case, $X$ would proceed to the next phase and broadcast its $\MName{Commit}$ message.
Hence, to prevent any resource contention, we designate only one {\em worker-thread} the task of processing all 
these messages.

When the input-thread receives a $\MName{Prepare}$ message, 
it enqueues that message in the {\em work-queue}. 
The worker-thread dequeues a message and verifies the signature on this message. 
If the verification is successful, then it records this message and
continues collecting $\MName{Prepare}$ messages 
corresponding to a $\MName{Pre-prepare}$ message until 
its count reaches $2f$.
Once it reaches this count, then it creates a $\MName{Commit}$ message, signs  
and broadcasts this message.
The worker-thread follows similar steps for a $\MName{Commit}$ message, 
except that it needs a total of $2f+1$ messages, and 
once it reaches this count, it informs the 
{\em execute-thread} to execute the client requests.

\subsection{Modeling a Backup Replica}
As a backup replica does not create batches of client requests, 
\expodb{} assigns it fewer threads.
When the input-thread at a backup replica receives a 
$\MName{Pre-prepare}$ message from the primary, then it 
enqueues it in the work-queue. 
The worker-thread at a backup dequeues a $\MName{Pre-prepare}$ message 
and checks if the message has a valid signature of the primary.
If this is the case, then the worker-thread creates a $\MName{Prepare}$ message, 
signs this message, and enqueues it in the queue for output-thread. 
Note that this $\MName{Prepare}$ message includes the digest from the $\MName{Pre-prepare}$ 
message and the sequence number suggested by the primary.
The output-thread broadcasts this $\MName{Prepare}$ message on the network.
Similar to the primary, each backup replica also collects $2f$ $\MName{Prepare}$ 
messages, creates and broadcasts a $\MName{Commit}$ message, collects $2f+1$ $\MName{Commit}$ messages, 
and informs the execute-thread.

\subsection{Out-of-Order Message Processing}
The key to the fast ordering of client requests is to allow ordering of multiple client requests to happen in parallel. 
\expodb{} supports parallel ordering of client requests, while ensuring a single common order across all the replicas.

\begin{example}\label{ex:out-of-order}
Say a client $C$ sends the primary replica $P$ first request $m_1$ and then request $m_2$. 
The input-thread at the primary $P$ would assign a sequence number $k$ to request $m_1$ and $k+1$ to request $m_2$.
However, as the batch-threads can work at varying speeds, so it is possible that the consensuses for requests $m_1$ 
and $m_2$ may either overlap, or some replica $R$ may receive $2f+1$ $\MName{Commit}$ messages for $m_2$ before $m_1$. 
\end{example}

In principle, Example~\ref{ex:out-of-order} seems like a challenge for a blockchain application,
as a replica may receive requests at sequence number $k+1, k+2, ...$ before it commits request at number $k$.
However, the property of out-of-order message processing is inherent in the design of most \bft{} protocols and 
is often overlooked.
 
Existing \bft{} protocols expect all the non-faulty replicas to act deterministic, that is, 
on identical inputs present identical outputs~\cite{pbft,zyzzyva,hotstuff}. 
Further, they only accept a request after they have a guarantee that a majority of other replicas 
have also accepted the same request. 
For example, in the \pbft{} protocol, say a backup replica $R$ receives a $\MName{Pre-Prepare}$ message for client request 
$m_1$ with sequence number $k$. 
This replica $R$ will not send a $\MName{Commit}$ message 
in support of the request $m_1$ 
until it receives $2f$ identical $\MName{Prepare}$ messages from distinct replicas in support of $m_1$.
Further, the replica $R$ will only execute request $m_1$ when it receives $2f+1$ $\MName{Commit}$ messages 
from distinct replicas.

In the case of out-of-order message processing, if a replica gets $2f+1$ $\MName{Commit}$ messages 
for a request with sequence number $k+1$ before the request with number $k$, it will {\em not execute} 
$(k+1)$-th request before $k$-th request. 
Hence, the execution of all the succeeding requests has to be kept on hold.
This ensures that the order of execution is identical across all the non-faulty replicas. 

Of course, the primary $\mathcal{P}$ could act malicious and could send all but the $k$-th request. 
To tackle such a scenario, \bft{} protocols already provide a primary-replacement 
(or {\em view-change}) algorithm~\cite{pbft,zyzzyva}.
The aim of the view-change algorithm is to deterministically replace the malicious primary $\mathcal{P}$ 
with a new primary $\mathcal{P}'$. 
It is the duty of this new primary $\mathcal{P}'$ to ensure all the replicas reach the common state 
otherwise it will also be replaced.
As \expodb{} uses existing \bft{} protocols, we skip presenting the details of existing view-change algorithm.

\subsection{Efficient Ordered Execution}
Although we parallelize consensus, we ensure execution happens in order. 
For instance, the requests $m_1$ and $m_2$ from Example~\ref{ex:out-of-order} are executed in sequence order, 
that is, $m_1$ is executed before $m_2$, irrespective of the order their consensuses completed.
At each replica, we dedicate a separate {\em execution-thread} to execute the requests. 
But, the key question remains: {\em how can we reduce the execution-thread's overhead of ordering.}

It is evident that the execution-thread has to wait for a notification from the worker-thread. 
In specific, we require the worker-thread to create an $\MName{Execute}$ message and 
place this message in the {\em appropriate} queue for the execution-thread.
This $\MName{Execute}$ message contains the identifier for the starting and ending transactions 
of a batch, which need to be executed.
Note that we associate a large set of queues with the execution-thread. 
To determine the number of required queues for the execution-thread, we use the 
parameter $QC$.
\begin{equation*}
QC = 2 \times Num\_Clients \times Num\_Req
\end{equation*}
Here, $Num\_Clients$ represent the total number of clients in the system, while $Num\_Req$ 
represents the maximum number of requests a client can send without waiting for any response.
We assume both of these parameters to be finite. 
Although $QC$ can be very large, the queues are logical. 
So, the space complexity remains almost the same as for a single queue.
But why is this practice advantageous?

Using this design the execute-thread can deterministically select the queue to dequeue.
%
If $k$ was the sequence number for last executed request, the execute-thread calculates 
$r = (k+1) ~\bmod ~QC$ and waits for an $\MName{Execute}$ message to be enqueued in its $r$-th queue.
This design is more efficient than having a single queue, as a single queue would have 
forced several dequeues and enqueues until finding the next request in order to execute.
%
Alternatively, we could have employed {\em hash-maps} but collision resistant hash functions 
are expensive to compute and verify~\cite{cryptobook}.

Once the execution is complete, the execution-thread creates a 
$\MName{Response}$ message and enqueues it in the queue for output-threads, to send to the client.
Note that ensuring execution happens in order provides a guarantee that 
a single common order is established across all the non-faulty replicas.

{\bf \em Block Generation.} 
It is at this stage where we require the execution thread to create a block representing 
this batch of requests.
As the execute-thread has access to the previous block in the chain, so it can easily hash this 
previous block and store this hash in the new block.
Note that this step provides another opportunity for parallelism where the 
execute-thread can delegate the task of creating a new block to another thread.

%
%
%
%
%
%

\subsection{Checkpointing}
We also require replicas to periodically generate and exchange {\em checkpoints}.
These checkpoints serve {\em two} purposes:
(1) Help a failed replica to update itself to the current state. 
(2) Facilitate cleaning of old requests, messages and blocks.
However, as checkpointing requires exchange of large messages, so we ensure it does not 
impact the throughput of the system.
\expodb{} deploys a separate {\em checkpoint-thread} at each replica to 
collect and process incoming $\MName{Checkpoint}$ messages. 
These checkpoint messages simply include all the blocks generated since the last checkpoint.
In specific, a $\MName{Checkpoint}$ message is sent only after a replica has executed $\Delta$ requests.
Once execute-thread completes executing a batch, it checks if 
the sequence number of the batch is a {\em multiple of} $\Delta$. 
If such is the case, it sends a $\MName{Checkpoint}$ message 
to all the replicas.
When a replica receives $2f+1$ identical $\MName{Checkpoint}$ messages from distinct replicas, 
then it marks the checkpoint and clears all the data before the previous checkpoint~\cite{pbft,zyzzyva}.

\subsection{Buffer Pool Management}
Until now, our description revolved around how a replica uses messages and transactions. 
In \expodb{}, we designed a {\em base class} that represents all the messages. 
To create a new message type, one has to simply inherit this base class and add required properties.
Although on delivery to the network, each message is simply a buffer of characters, 
this typed representation helps us to easily manipulate the required properties.
Similarly, we have designed a {\em base class} to represent all client transactions.
An object of this transaction class includes: transaction identifier, 
client identifier, and transaction data, among many other properties. 

When a message arrives in the system, a replica needs to allocate ({\tt malloc}) 
space for that messages.
Similarly, when a replica receives a client request, it needs to allocate corresponding 
transaction objects.
When the lifetime of a message ends (or a new checkpoint is established), then the memory 
occupied by that message (or transactions object) needs to be released ({\tt free}).
To avoid such frequent allocations and de-allocations, we adopt the standard practice 
of maintaining a set of {\em buffer pools}.
At the system initialization stage, we create a large number of empty objects representing 
the messages and transactions. 
So instead of doing a {\tt malloc}, these objects are extracted from their respective pools and 
are placed back in the pool during the {\tt free} operation.

\section{Experimental Analysis}
\label{s:eval}

We now analyze how various parameters affect the throughput and latency of a 
Permissioned Blockchain (henceforth abbreviated as \pbc{}) system.
For the purpose of this study we use our \expodb{} fabric. 
Although \expodb{} can employ any \bft{} consensus protocol, 
we use the simple \pbft{} protocol to ensure the system design remains as our key focus.
To ensure a holistic evaluation, we attempt to answer the following questions:
\begin{enumerate}[nosep,label=(Q\arabic*),ref={Q\arabic*}]
\item \label{q:0}Can a well-crafted system based on a classical \bft{} protocol outperform a modern protocol?
\item \label{q:1}How much gains in throughput (and latency) can a \pbc{} achieve from pipelining and threading?
\item \label{q:2}Can pipelining help a \pbc{} become more scalable? 
\item \label{q:batch}What impact does batching of requests has on a \pbc{}?
\item \label{q:op}Do multi-operation requests impact the throughput and latency of a \pbc{}?
\item \label{q:msize}How increasing the message size impacts a \pbc{}?
\item \label{q:crypto}What effect do different types of cryptographic signature schemes have on the throughput of a \pbc{}?
\item \label{q:mem}How does a \pbc{} fare with in-memory storage versus a storage provided by a standard database?
\item \label{q:clients}Can an increased number of clients impact the latency of a \pbc{}, while its throughput remains 
unaffected?
\item \label{q:cores}Can a \pbc{} sustain high throughput on a setup having fewer number of cores?
\item \label{q:fail}How impactful are replica failures for a \pbc{}?
\end{enumerate}

\subsection{Evaluation Setup}
We employ Google Cloud infrastructure at Iowa region to deploy our \expodb{}. 
For replicas, we use {\tt c2} machines with an $8$-core Intel Xeon Cascade Lake CPU running at $3.8$GHz
and having $16$GB memory, 
while for clients we use {\tt c2} $4$-core machines.
We run each experiment for $180$ seconds, and
collect results over {\em three} runs to average out any noise.

We use YCSB~\cite{blockbench,ycsb} for generating workload for client requests.
For creating a request, each client indexes a YCSB table with an active set of $600$K records.
In our evaluation, we require client requests to contain only write accesses, as a majority of 
blockchain requests are updates to the existing data.  
During the initialization phase, we ensure each replica has an identical copy of the table. 
Each client YCSB request is generated from a uniform Zipfian distribution.

Unless {\em explicitly} stated otherwise, we use the following setup:
We invoke up to $80$K clients on $4$ machines and run consensus among $16$ replicas.
We employ batching to create batches of $100$ requests.
For communication among replicas and clients we employ digital signatures based on ED25519, and 
for communication among replicas we use a combination of CMAC and AES~\cite{cryptobook}.
At each replica, we permit one worker-thread, one execute-thread and two batch-threads

\begin{figure}[t]
    \centering
    \begin{subfigure}[t]{0.4\columnwidth}
	\centering
	\includegraphics[width=\textwidth]{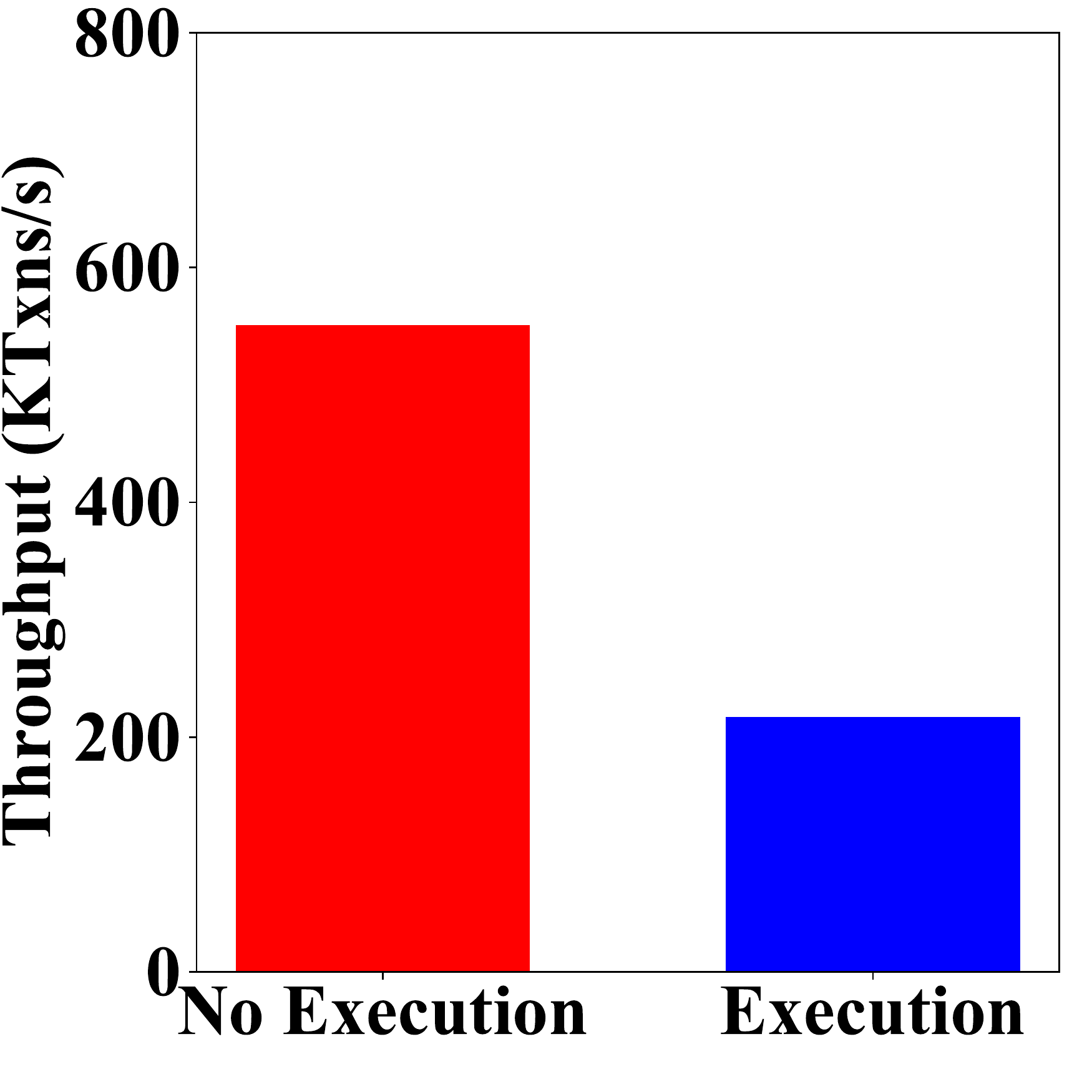}
	\caption{\small System throughput.}
	\label{fig:upper-tput}
    \end{subfigure}
    \begin{subfigure}[t]{0.4\columnwidth}
	\centering
	\includegraphics[width=\textwidth]{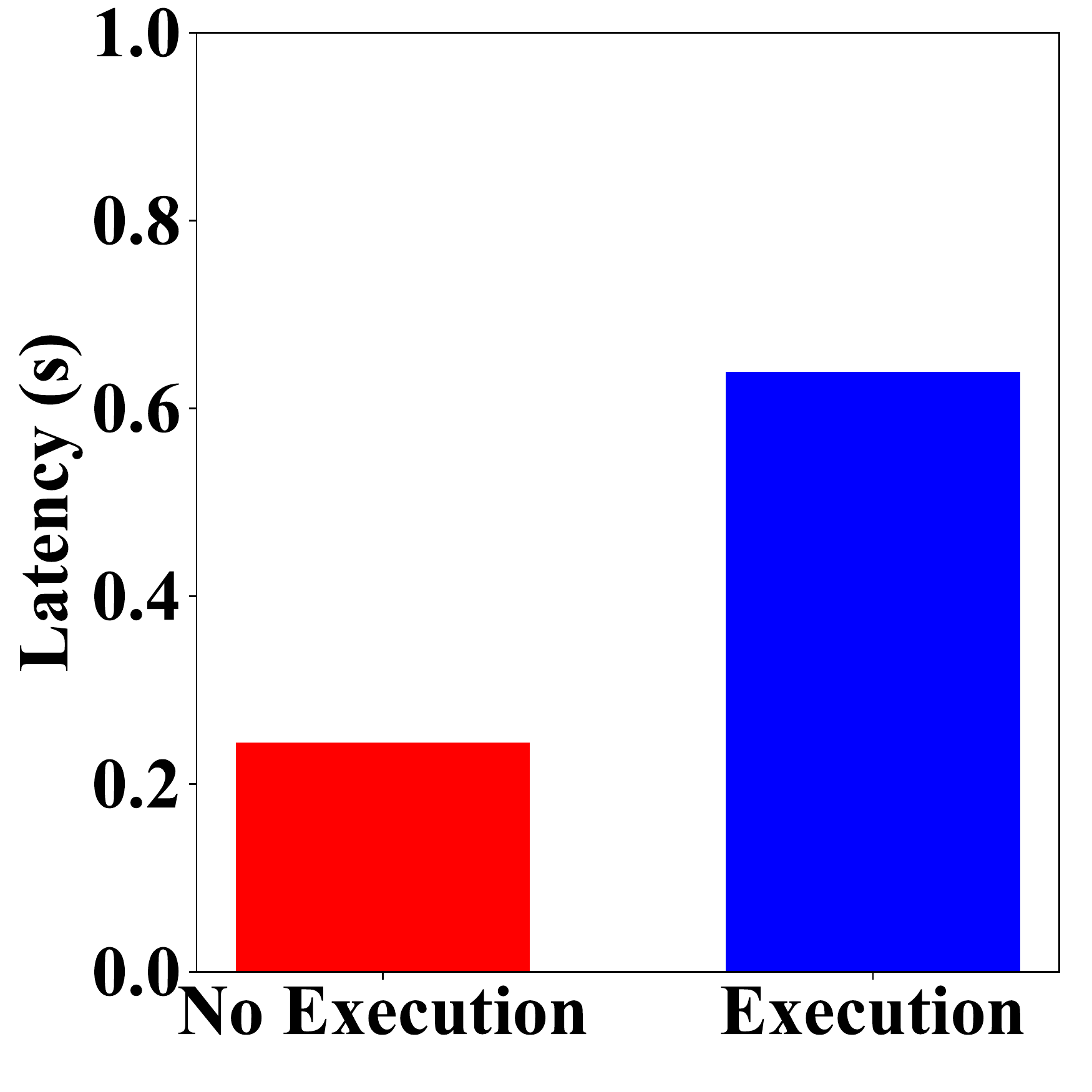}
	\caption{\small Latency.}
	\label{fig:upper-lat}
    \end{subfigure}
      \caption{Upper bound measurements: (i) Primary responds back to the client 
	without Execution, and (ii) Primary executes the request and replies to the client.}
\end{figure}

\subsection{Effect of Threading and Pipelining}
In this section, we analyze and answer questions~\ref{q:0} to~\ref{q:2}.
For this study, we vary the system parameters in two dimensions:
(i) We increase the number of replicas participating in the consensus from $4$ to $32$. 
(ii) We expand the pipeline and gradually balance the load among parallel threads.

We first try to gauge the upper bound performance of our system.  
In Figures~\ref{fig:upper-tput} and~\ref{fig:upper-lat}, we measure the maximum throughput and latency 
a system can achieve, when there is no communication among the replicas or any consensus protocol.
We use the term {\em No Execution} to refer to the case where all the clients send their requests to 
the primary replica and primary simply responds back to the client. We count every query responded back 
in the system throughput.
We use the term {\em Execution} to refer to the case where the primary replica executes each query 
before responding back to the client.
In both of these experiments, we allowed two threads to work independently at the primary replica, 
that is, no ordering is maintained.
Clearly, the system can attain high throughputs (up to $500K$ txns/s) and has low latency 
(up to $0.25$s). 

\begin{figure}[t]
    \begin{subfigure}[t]{0.5\textwidth}
	\centering
	\includegraphics[width=0.9\columnwidth]{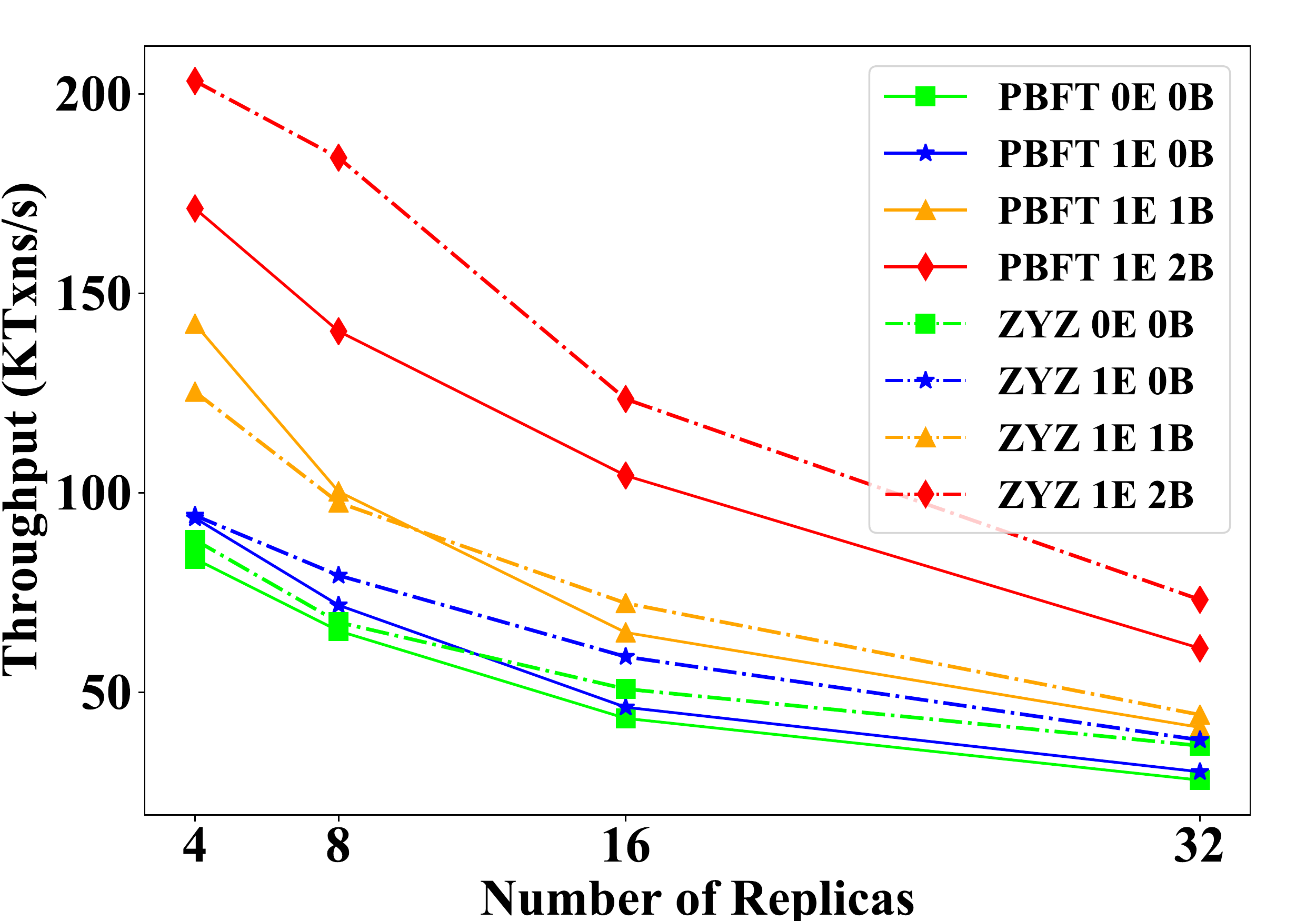}
	\caption{\small System throughput.}
	\label{fig:tput-pipe}
    \end{subfigure}
    \begin{subfigure}[t]{0.5\textwidth}
	\centering
	\includegraphics[width=0.9\columnwidth]{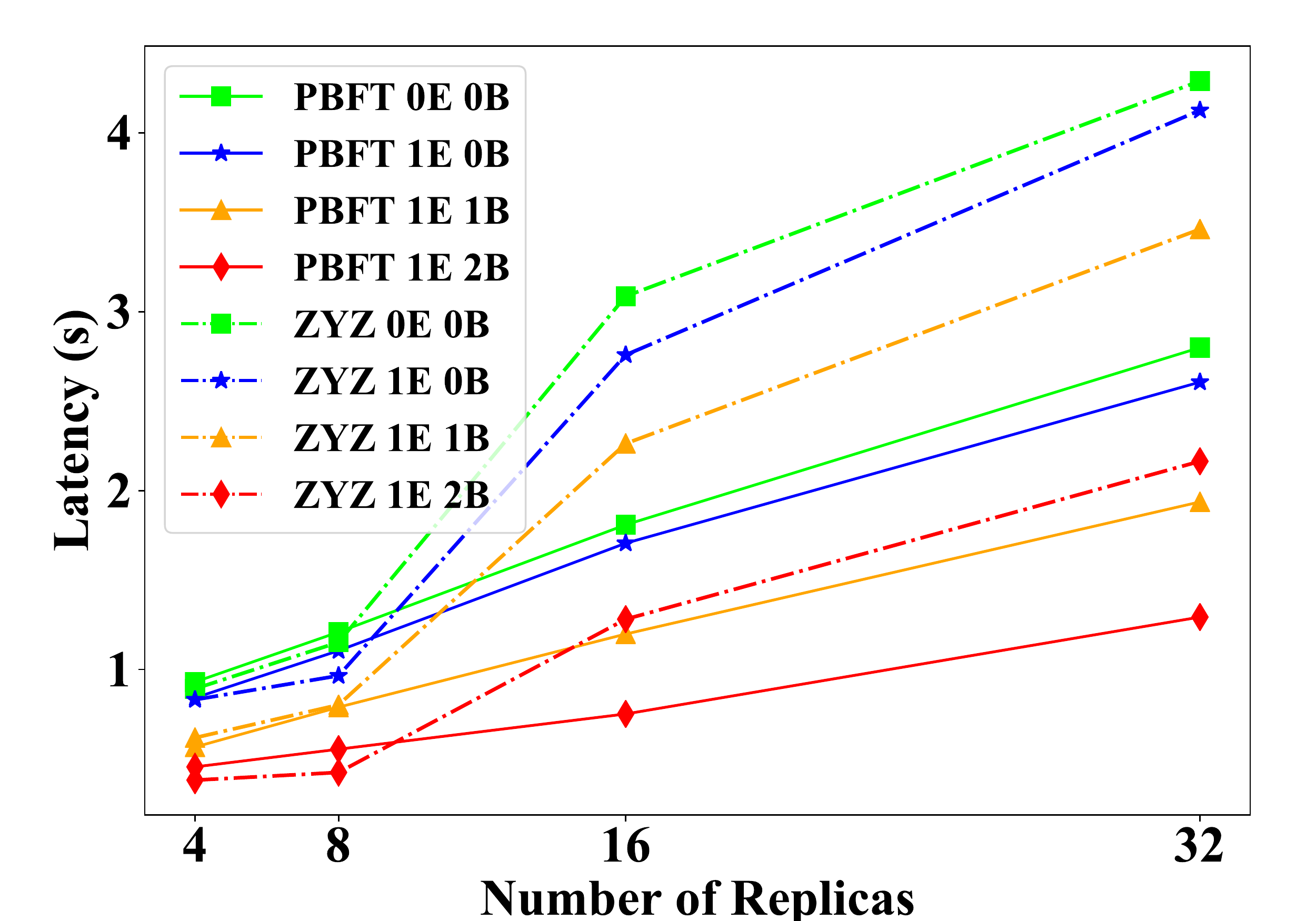}
	\caption{\small Latency.}
	\label{fig:lat-pipe}
    \end{subfigure}
      \caption{System throughput and latency, on varying the number of replicas participating 
	in the consensus. Here, E denotes number of execution-threads, while B denotes batch-threads.}
\end{figure}

Next, we take two consensus protocols: \pbft{} and \zyzzyva{}, and we ensure that at least $3f+1$ replicas 
are participating in the consensus.
We gradually move our system towards the architecture of Figure~\ref{fig:prim-pipe}.
In Figures~\ref{fig:tput-pipe} and~\ref{fig:lat-pipe}, we show the effects of this gradual increase.
We denote the number of execution-threads with symbol E, and batch-threads with symbol B.
For all these experiments, we used only {\em one} worker-thread. 
The key intuition behind these plots is to continue expanding the stages of pipeline and 
the number of threads, until system can no longer increase its throughput. 
In this manner, it would be easy to observe design choices that could make even \pbft{} outperform 
\zyzzyva{}, that is, benefits of a {\em well-crafted implementation}.

\begin{figure*}
    \centering
    \begin{subfigure}[t]{0.8\textwidth}
	\centering
	\includegraphics[width=\columnwidth]{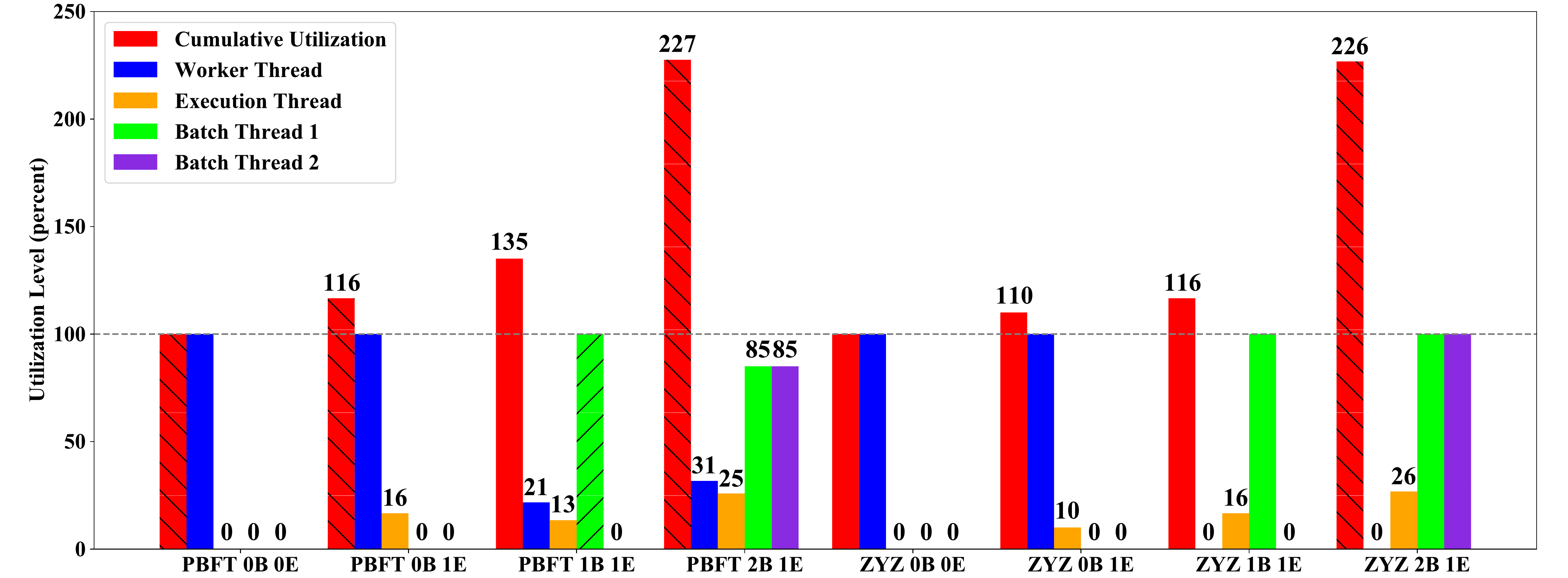}
	\caption{\small Primary Replica.}
	\label{fig:prim-thd-dist}
    \end{subfigure}

    \begin{subfigure}[t]{0.8\textwidth}
	\centering
	\includegraphics[width=\columnwidth]{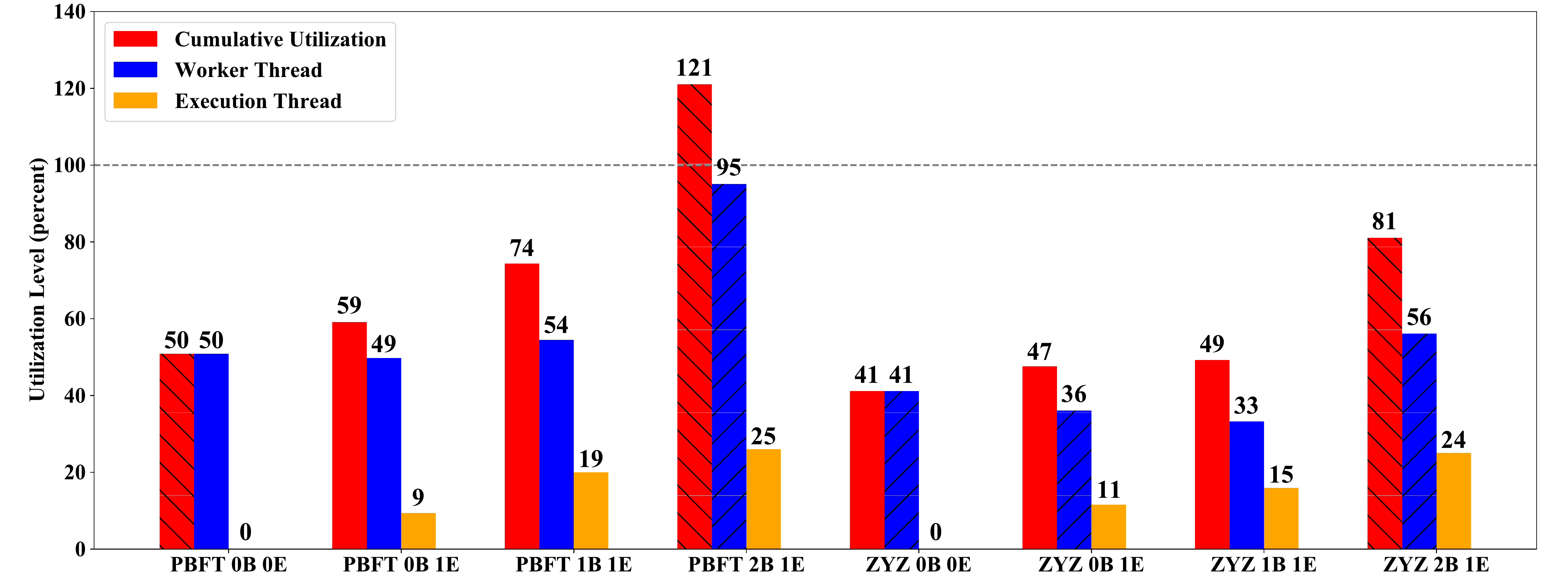}
	\caption{\small Backup Replica.}
	\label{fig:non-prim-thd-dist}
    \end{subfigure}
      \caption{Utilization level of different threads at a replica. The mean is at $100\%$, which implies the thread is completely utilized.}
\end{figure*}


\begin{figure}[t]
	\centering
    \begin{subfigure}[t]{0.6\columnwidth}
	\includegraphics[width=\textwidth]{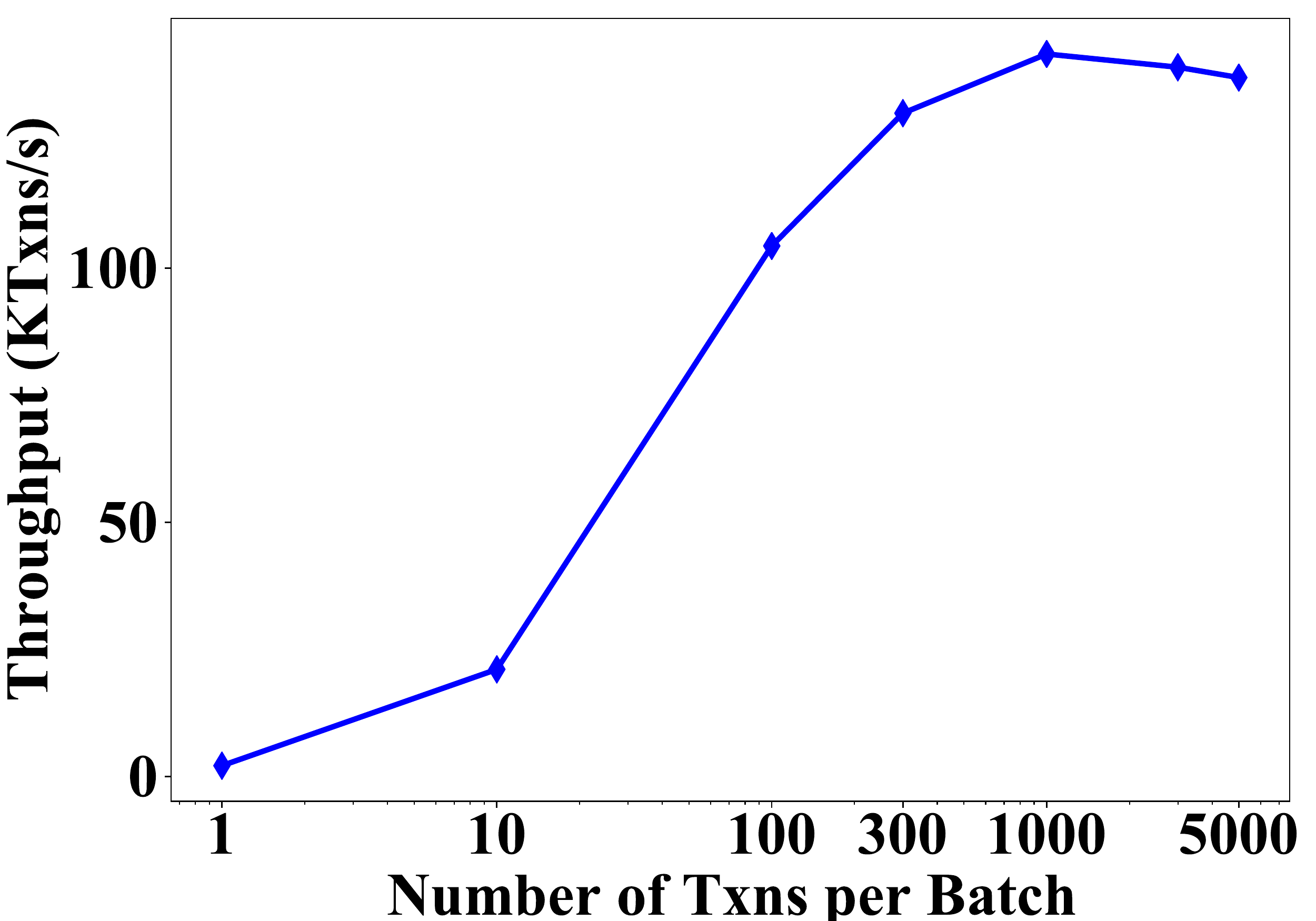}
	\caption{\small System throughput.}
	\label{fig:tput-batch}
    \end{subfigure}
    \begin{subfigure}[t]{0.6\columnwidth}
	\centering
	\includegraphics[width=\textwidth]{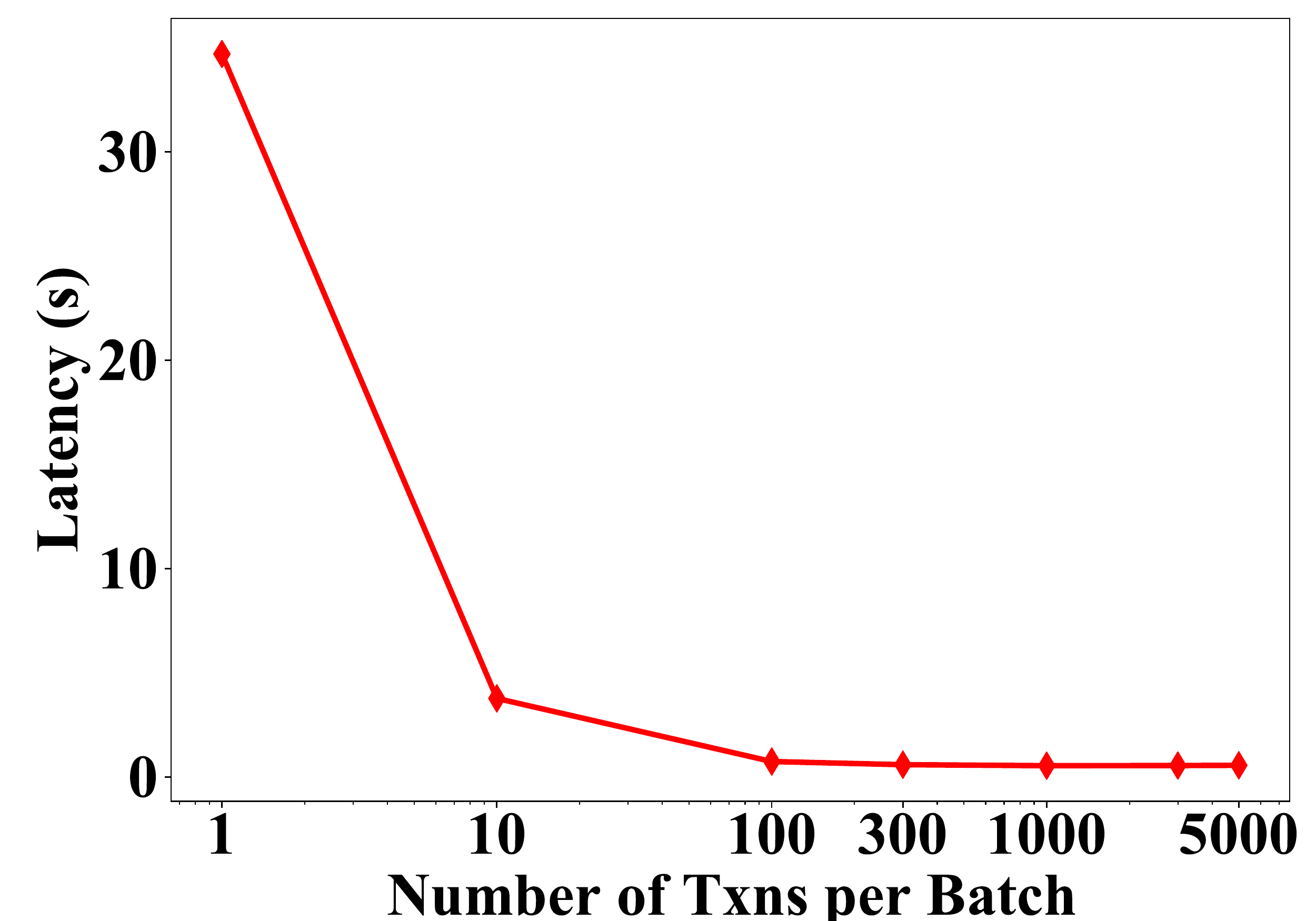}
	\caption{\small Latency.}
	\label{fig:lat-batch}
    \end{subfigure}
    \caption{System throughput and latency on varying the number of transactions per batch.
	In this experiment, $16$ replicas participate in consensus.}
    \label{fig:batch}
\end{figure}

On close observation of Figure~\ref{fig:tput-pipe}, we can trivially highlight the benefits 
of a good implementation. 
Further, these plots help to confirm our intuition that a multi-threaded pipelined architecture for a \pbc{} 
outperforms a single-threaded design. 
This is the key reason why our design of \expodb{} employs {\em one} execution-thread and
{\em two} batch-threads apart from a single worker-thread.
 
Next, we explain our methodology for gradual changes. 
We first modified \expodb{} to ensure there are no additional threads for 
execution and batching, that is, all tasks are done by one worker-thread (0E 0B). 
On scaling this system we realized that this worker-thread was getting fully utilized. 
Hence, we partially divide the load by having an execute-thread (1E 0B).
However, we again observed that the worker-thread at the primary was getting completely utilized. 
So we had an opportunity to introduce a separate thread to create batches (1E 1B). 
Although worker-thread was no longer saturating, the batch-thread was overloaded with the task of creating 
batches. 
Hence, we further divided the task of batching among multiple batch-threads (1E 2B) and 
ensured none of the batch-threads were fully utilized.
Figures~\ref{fig:prim-thd-dist} and~\ref{fig:non-prim-thd-dist} show the utilization level 
for different threads at a replica.
In this figure, we mark $100\%$ as the maximum utilization for any thread. 
Using the bar for {\em cumulative utilization}, we show a summation of the utilization
for all the threads, for any experiment.
Note that for \pbft{} 1E 2B, the worker-thread at the backup replicas have started to saturate. 
But, as the architecture at the non-primary is following our design, so we split no further.

It can be observed that if \pbft{} is given benefit of \expodb's standard pipeline (1E 2B), 
then it can attain higher throughput than all but one \zyzzyva{} implementations. 
The only \zyzzyva{} implementation (1E 2B) that outperforms \pbft{} is the one that employs 
\expodb's standard threaded-pipeline.
Further, even the simpler implementation for \pbft{} (1E 1B) attains higher throughput than 
\zyzzyva's 0E 0B and 1E 0B implementations.

We have always stated that the design of \expodb{} is independent of the underlying consensus protocol. 
This can be observed from the fact that when \zyzzyva{} is given \expodb's standard pipeline, then 
it can achieve throughput of $200$K txns/s. 
Note that in majority of the settings \pbft{} incurs less latency than \zyzzyva{}. 
This is an effect of \zyzzyva's algorithm, which requires the client to wait for replies 
from all the $n$ replicas, where for \pbft{} the client only needs $f+1$ responses.
To {\bf \em summarize}: 
(i) \pbft's throughput (latency) increases (reduces) by $1.39\times$ ($58.4\%$)  
on moving from 0E 0B setup to 1E 2B. 
(ii) \zyzzyva's throughput (latency) increases (reduces) by $1.72\times$ ($63.19\%$) 
on moving from 0E 0B setup to 1E 2B.
(iii) Throughput gains up to $1.07\times$ are possible on running \pbft{} on an efficient setup,
in comparison to basic setups for \zyzzyva{}.

\subsection{Effect of Transaction Batching}
We now try to answer question~\ref{q:batch} by studying how batching the client transactions 
impacts the throughput and latency of a \pbc{}. 
For this study, we increase the 
size of a batch from $1$ to $5000$.

Using Figures~\ref{fig:tput-batch} and~\ref{fig:lat-batch}, we observe that as the number of 
transactions in a batch increases, the throughput increases until a limit (at $1000$) and then starts decreasing (at $3000$). 
At smaller batches, more consensuses are taking place, and hence communication impacts the system throughput. 
Hence, larger batches help reduce the consensuses.
However, when the transactions in a batch are increased further, 
then the size of the resulting message and the time taken to create a batch by a batch-thread,
reduces the system throughput.
Hence, any \pbc{} needs to find an optimal number of client transactions that it can batch.
To {\bf \em summarize:} batching can increase throughput by up to $66\times$ and 
reduce latency by up to $98.4\%$.


\begin{figure}[t]
    \begin{subfigure}[t]{\columnwidth}
	\centering
	\includegraphics[width=\columnwidth]{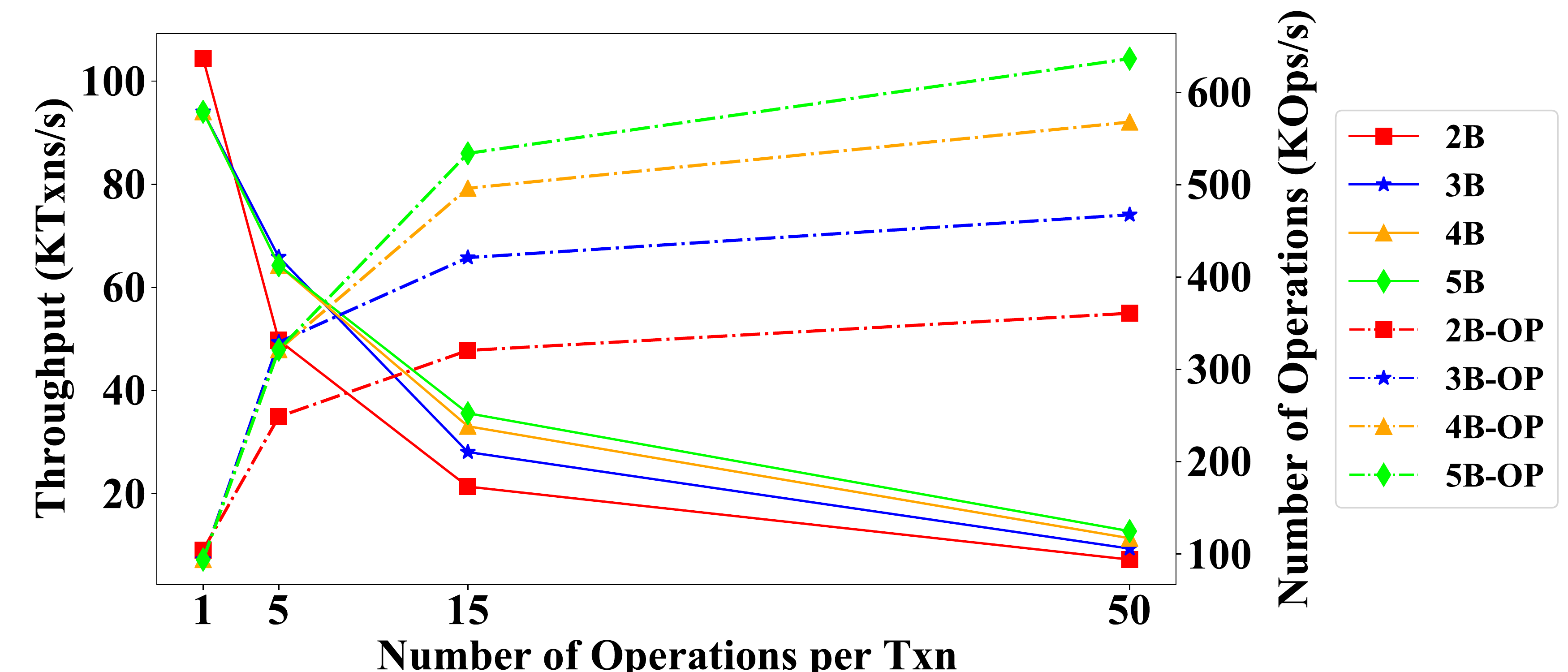}
	\caption{\small System throughput.}
	\label{fig:tput-ops}
    \end{subfigure}
    \begin{subfigure}[t]{\columnwidth}
	\centering
	\includegraphics[width=\columnwidth]{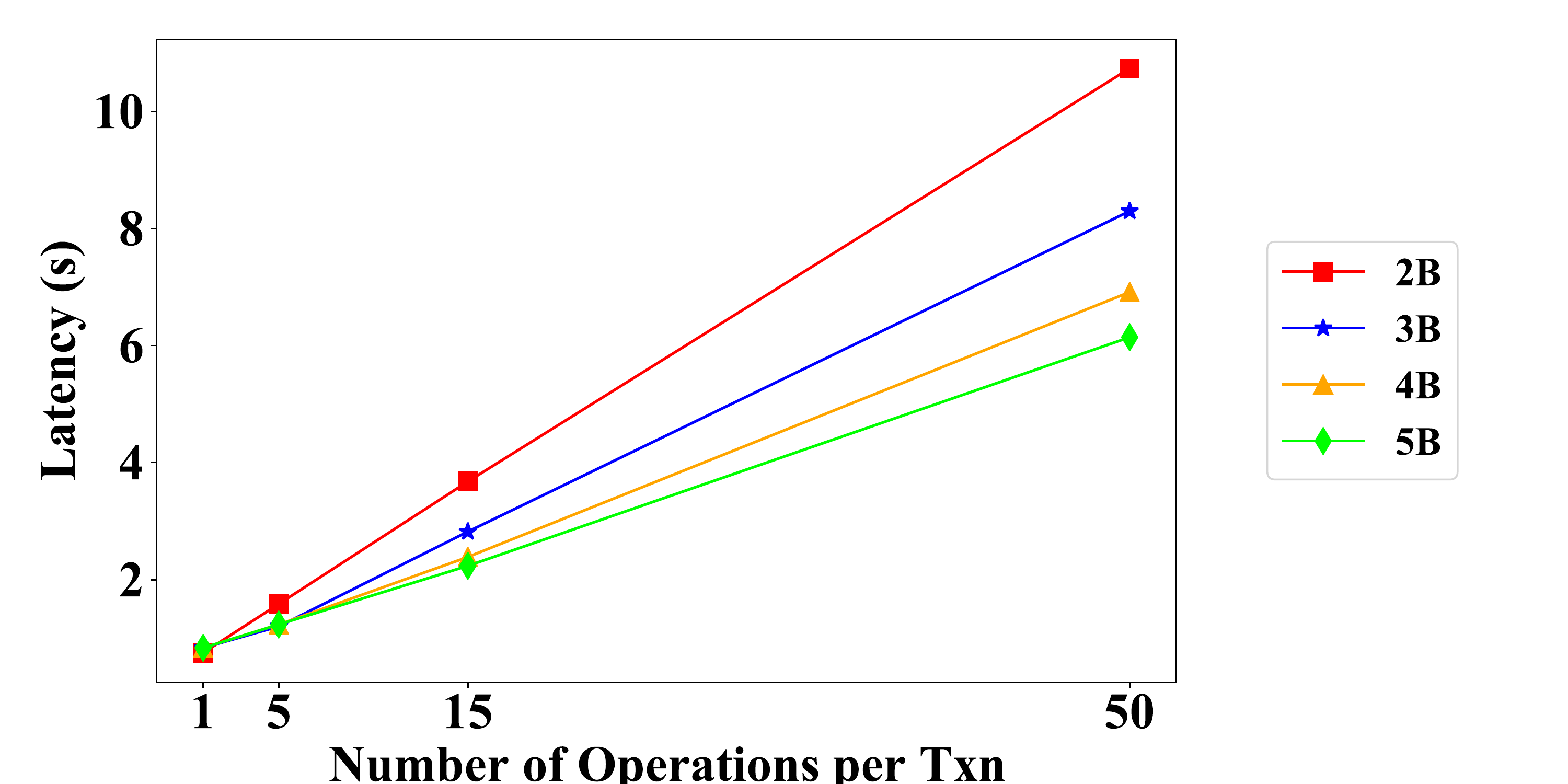}
	\caption{\small Latency.}
	\label{fig:lat-ops}
    \end{subfigure}
      \caption{System throughput and latency on varying the number of operations per transaction. Here, B denotes the number of batch-threads used in the experiment.}
\end{figure}

\subsection{Effect of Multi-Operation Transactions}
We now answer question~\ref{q:op}, that is, 
understand how multi-operation transactions affect the throughput of a system?
In Figures~\ref{fig:tput-ops} and~\ref{fig:lat-ops}, we increase the number of 
operations per transaction from $1$ to $50$. 
Further, we increase the number of batch-threads from $2$ to $5$, while having one worker-thread and 
one execute-thread.
Although multi-operation transactions are common, 
prior works do not provide any discussion on such transactions.
Notice that these experiments are orthogonal counterparts of the experiments in the previous section.

It is evident from these figures that on increasing the number of operations per transaction, the system throughput 
decreases. 
This decrease is a consequence of batch-threads getting saturated as they perform task 
of batching and allocating resources for transaction. 
Hence, we ran several experiments with distinct counts for batch-threads. 
An increase in the number of batch-threads helps the system to increase its throughput, but the 
gap reduces significantly after the transaction becomes too large (at $50$ operations).
Similarly, more batch-threads help to  decrease the latency incurred by the system.

Alternatively, we also measure the total number of operations completed in each experiment. 
Notice that if we base the throughput on the number of operations executed per second, then the 
trend has completely reversed.
Indeed, this makes sense as in fewer rounds of consensus, more operations have been executed.
To {\bf \em summarize:}
multi-operation transactions can cause a decrease of $93\%$ in throughput and 
an increase of $13.29\times$ in latency, on the two batch-threads setup.
An increase in batch-threads from two to five increases throughputs up to $66\%$ and 
reduces latencies up to $39\%$.


\begin{figure}[t]
    \centering
    \begin{subfigure}[t]{0.6\columnwidth}
	\centering
	\includegraphics[width=\textwidth]{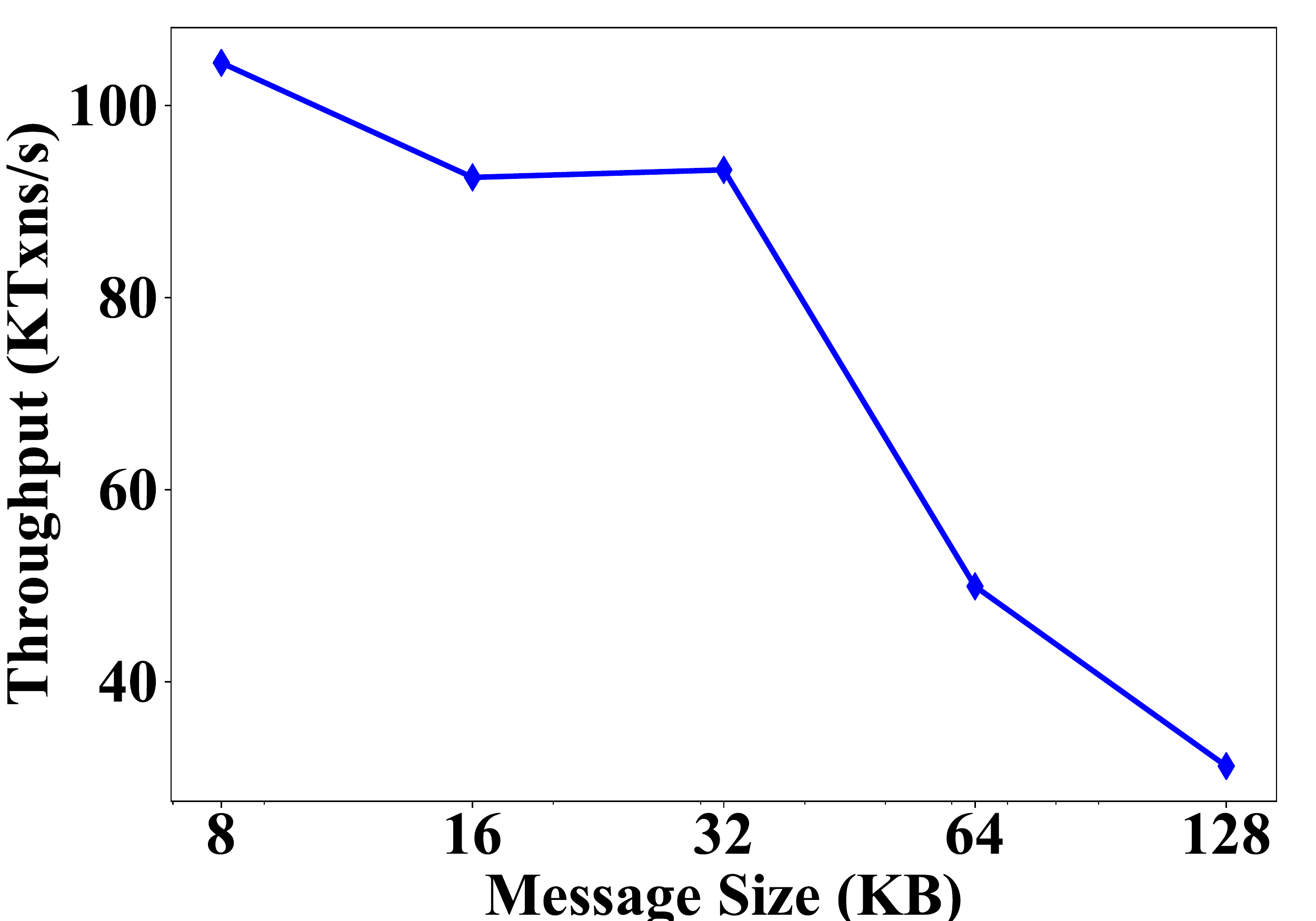}
	\caption{\small System throughput.}
	\label{fig:msize-tput}
    \end{subfigure}
    \begin{subfigure}[t]{0.6\columnwidth}
	\centering
	\includegraphics[width=\textwidth]{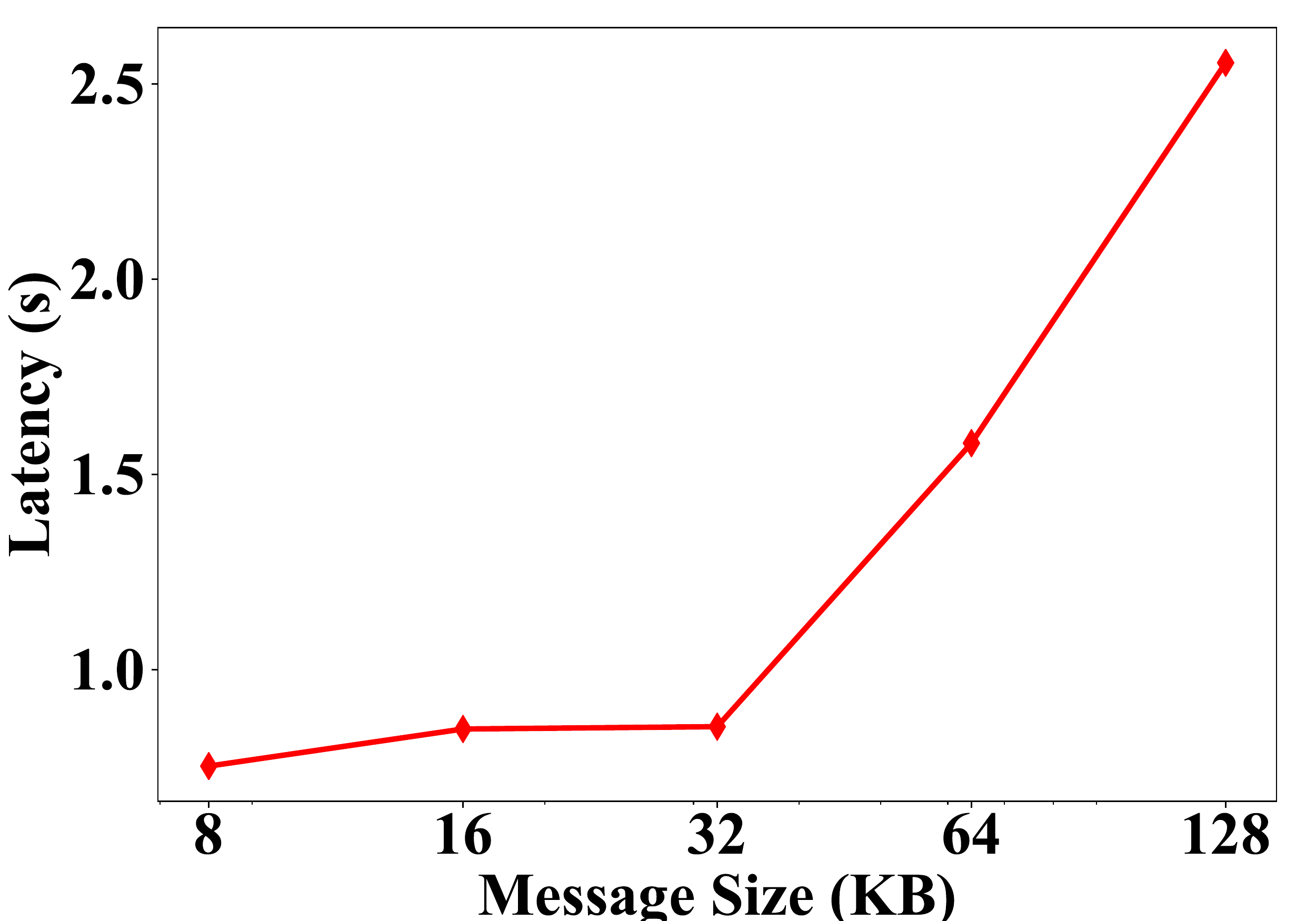}
	\caption{\small Latency.}
	\label{fig:msize-lat}
    \end{subfigure}
      \caption{System throughput and latency on varying the message size.
Here, $16$ replicas participate in consensus.}
\end{figure}


\begin{figure}[t]
    \centering
    \begin{subfigure}[t]{0.58\columnwidth}
	\centering
	\includegraphics[width=\textwidth]{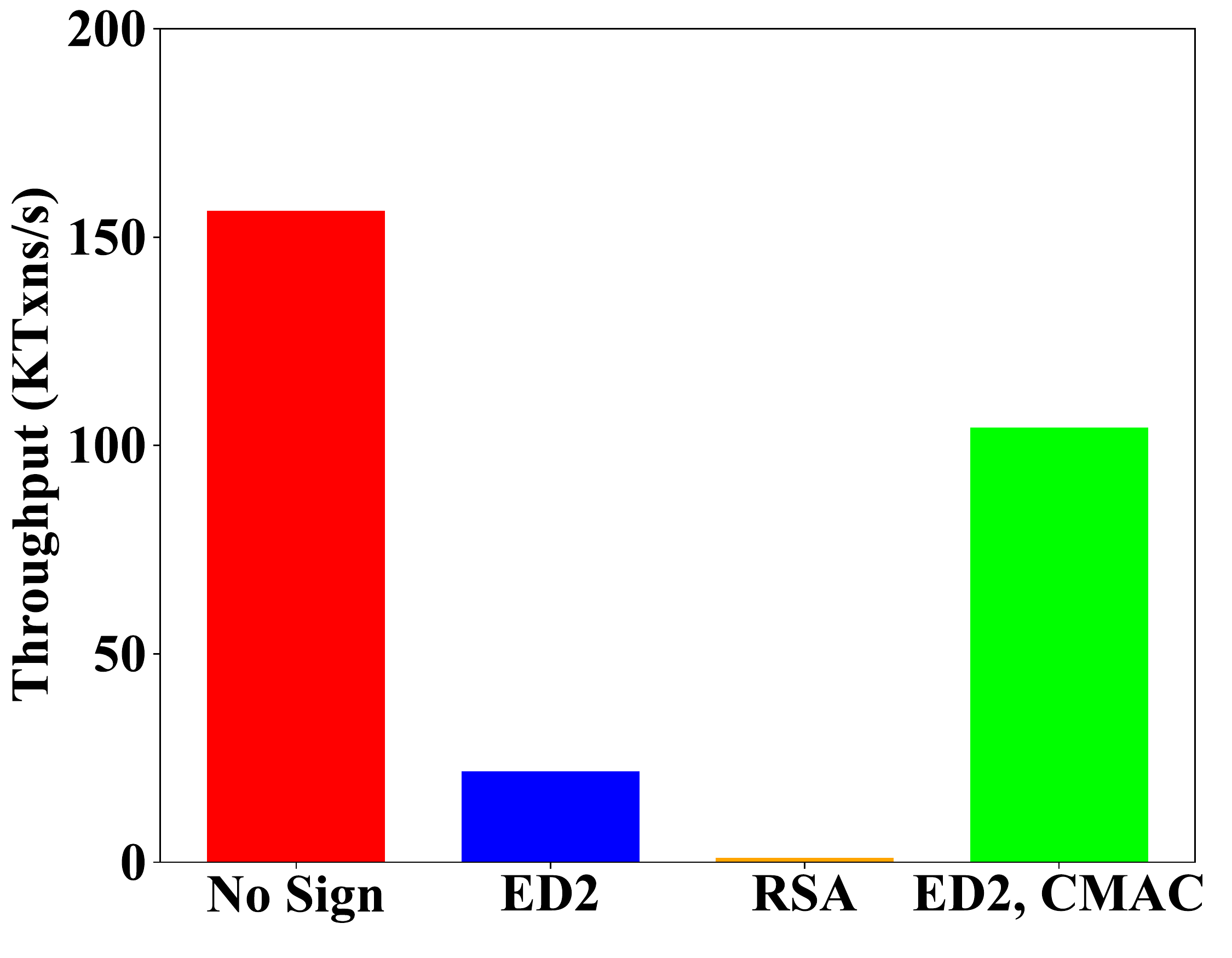}
	\caption{\small System throughput.}
	\label{fig:crypto-tput}
    \end{subfigure}
    \begin{subfigure}[t]{0.58\columnwidth}
	\centering
	\includegraphics[width=\textwidth]{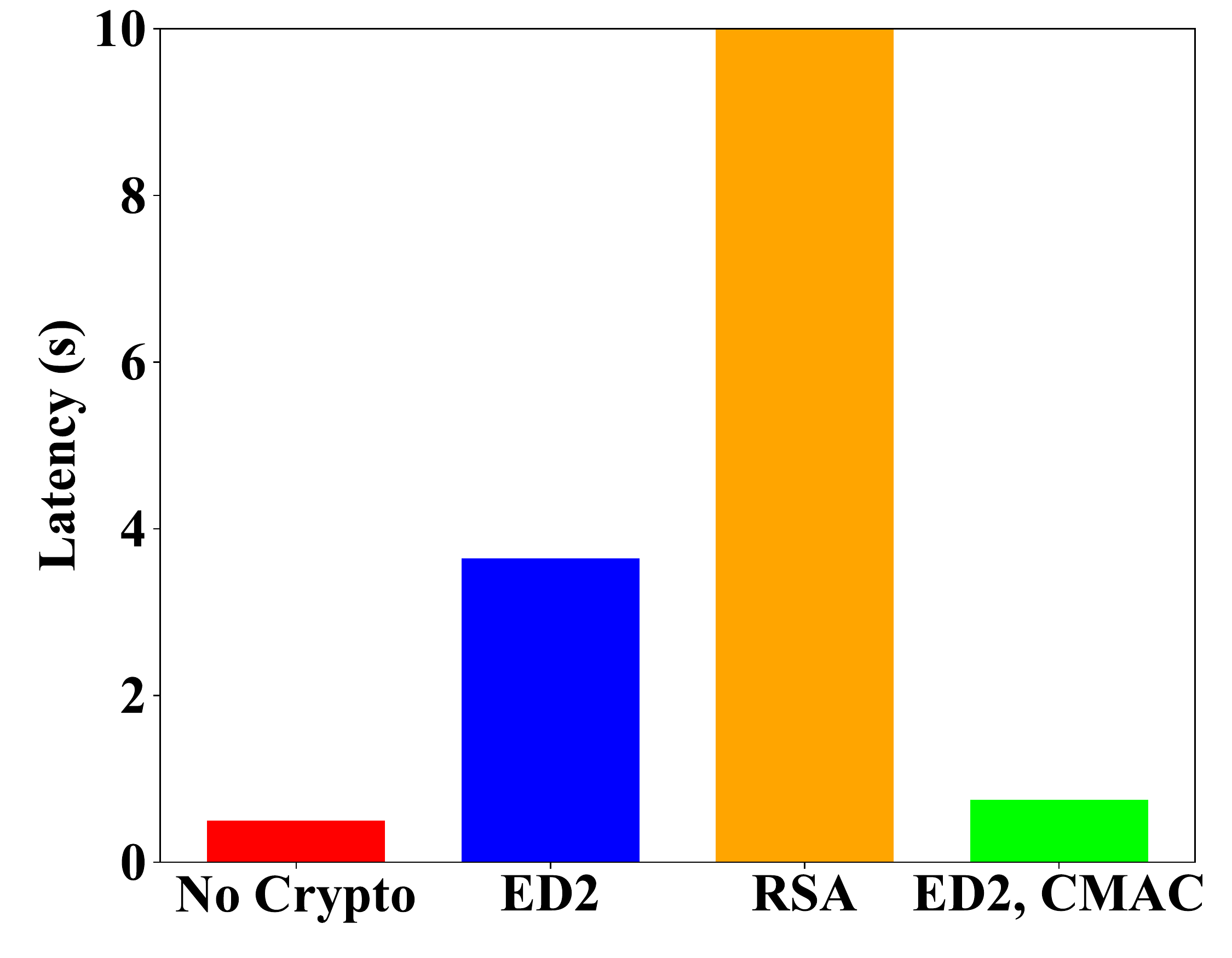}
	\caption{\small Latency.}
	\label{fig:crypto-lat}
    \end{subfigure}
      \caption{System throughput and latency with different signature schemes.
Here, $16$ replicas participate in consensus.}
\end{figure}

\subsection{Effect of Message Size}
We now try to answer question~\ref{q:msize} by increasing the size of the $\MName{Pre-prepare}$ 
message in each consensus.
The key intuition behind this experiment is to gauge how well a \pbc{} system performs when the requests sent 
by a client are large. 
Although each batch includes only $100$ client transactions, individually, these requests can be large.
Hence, these experiments are aimed at exploiting a different system parameter than the plots of Figure~\ref{fig:batch}.

In Figures~\ref{fig:msize-tput} and~\ref{fig:msize-lat}, we study the variation in throughput and latency 
by increasing the size of a $\MName{Pre-prepare}$ message. 
We do this by adding a payload to each message, which includes a set of integers ($8$byte each).
The cardinality of this set is kept equal to the desired message size.

It is evident from these plots that as the message size increases, there is a decrease in the system throughput 
and an increase in the latency incurred by the client. 
This is a result of network bandwidth becoming a limitation, due to which it takes extra time to push more data onto the network. 
Hence, in this experiment, the system reaches a network bound before any thread can hit its computational bound. 
This leads to all the threads being idle.
To {\bf \em summarize:}
On moving from $8$KB to $64$KB messages, there is a $52\%$ decrease in throughput and 
$1.09\times$ increase in latency.

\subsection{Effect of Cryptographic Signatures}
In this section, we answer question~\ref{q:crypto} by studying the impact of different 
cryptographic signature schemes. 
The key intuition behind these experiments is to determine which signing scheme helps 
a \pbc{} achieve the highest throughput while preventing byzantine attacks.
For this purpose, we run four different experiments to measure the system throughput and 
latency when:
(i) no signature scheme is used, 
(ii) everyone uses digital signatures based on ED25519, 
(iii) everyone uses digital signatures based on RSA, and
(iv) all replicas use CMAC+AES for signing, while clients sign their message using ED25519.

Figures~\ref{fig:crypto-tput} and~\ref{fig:crypto-lat} help us to illustrate the 
throughput attained and latency incurred by \expodb{} for different configurations.
It is evident that \expodb{} attains maximum throughput when no signatures are employed. 
However, such a system does not fulfill the minimal requirements of a permissioned 
blockchain system.
Further, using just digital signatures for signing messages is not exactly the best practice. 
An optimal configuration can require clients to sign their messages using digital signatures, 
while replicas can communicate using MACs.
To {\bf \em summarize:} 
(i) use of cryptography reduces throughput by at least $49\%$ and increases latency by $33\%$.
(ii) choosing RSA over CMAC, ED25519 combination would increase latency by $125\times$.


\begin{figure}[t]
    \centering
    \begin{subfigure}[t]{0.45\columnwidth}
	\centering
	\includegraphics[width=\textwidth]{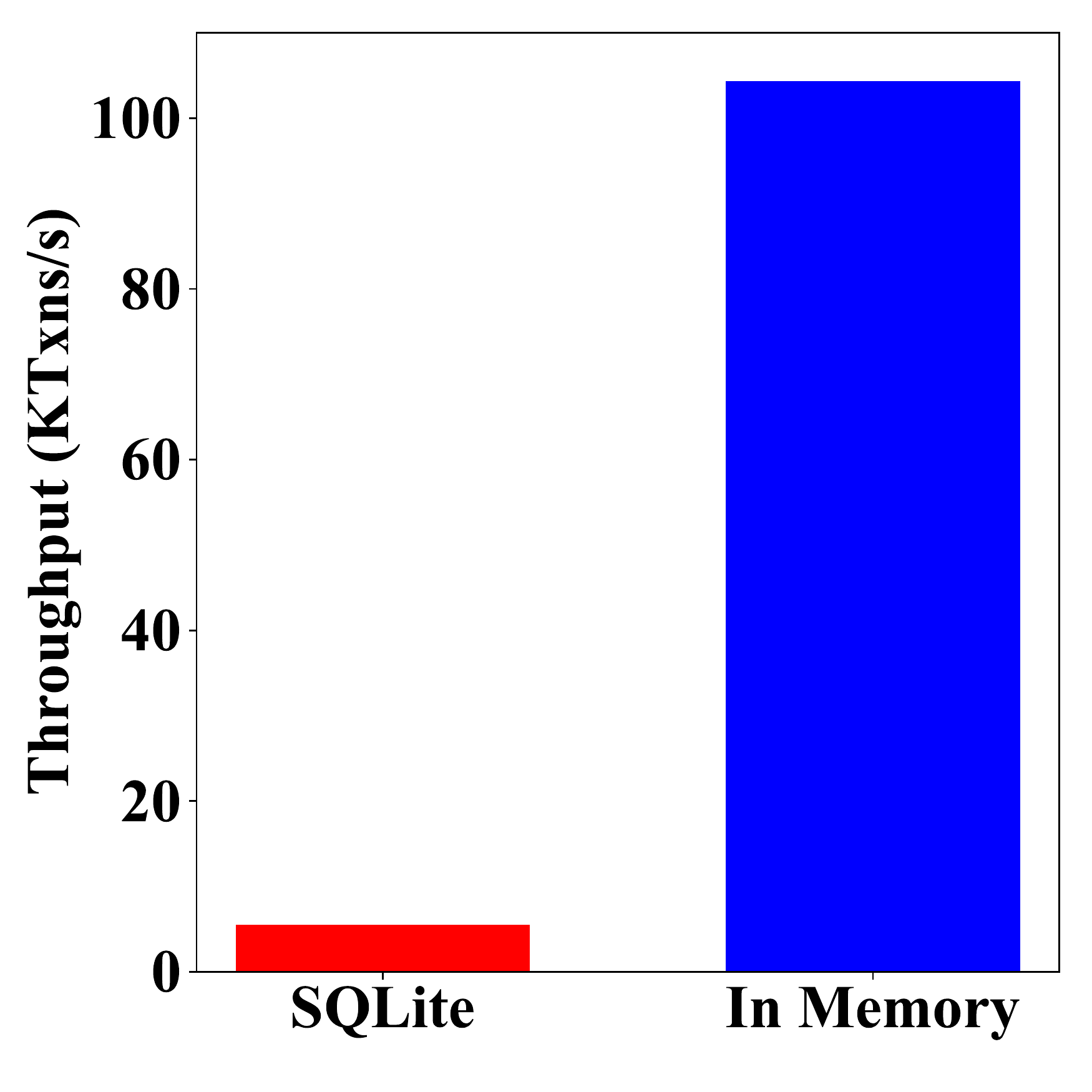}
	\caption{\small System throughput.}
	\label{fig:mem-tput}
    \end{subfigure}
    \begin{subfigure}[t]{0.45\columnwidth}
	\centering
	\includegraphics[width=\textwidth]{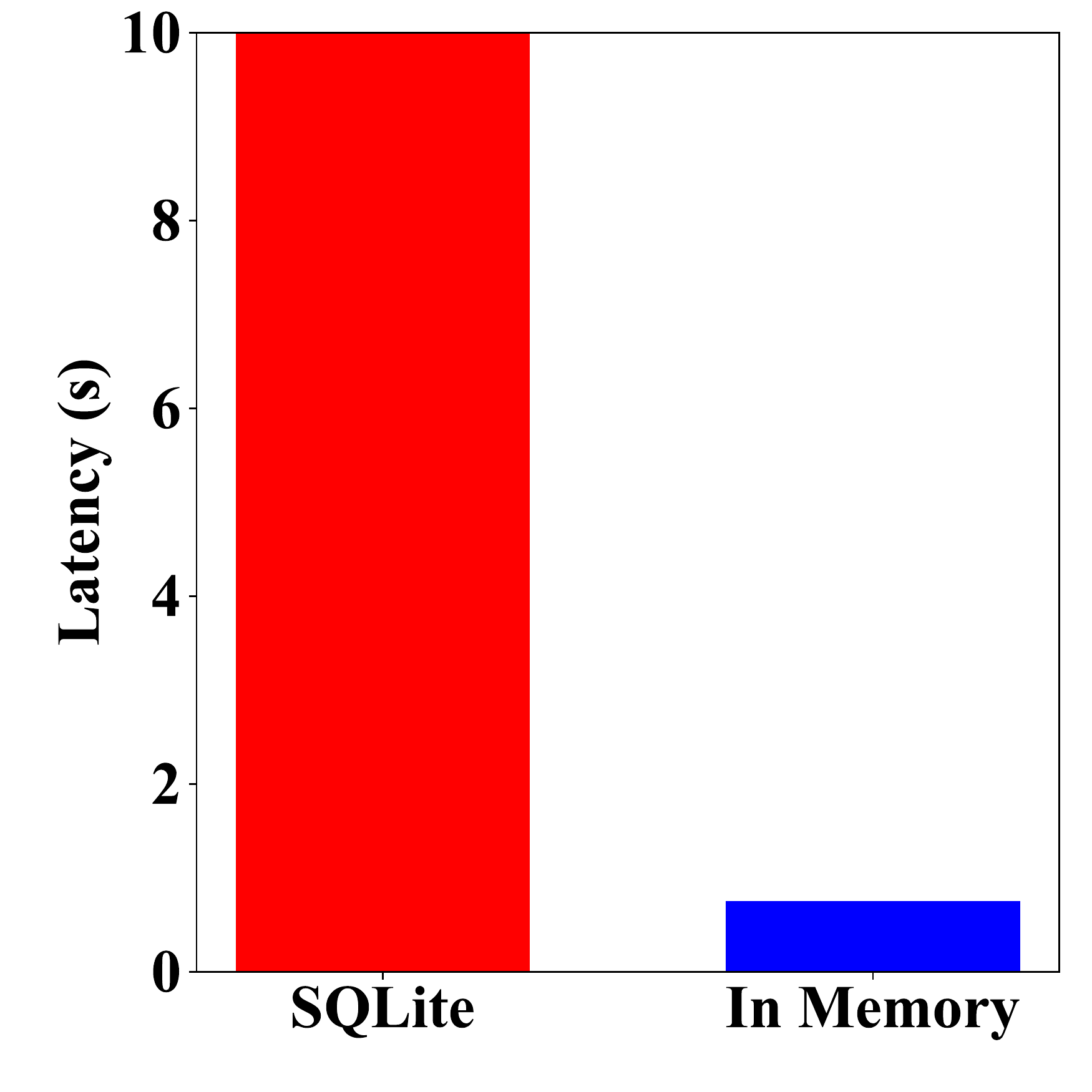}
	\caption{\small Latency.}
	\label{fig:mem-lat}
    \end{subfigure}
      \caption{System throughput and latency for in-memory storage vs. off-memory storage.
	Here, $16$ replicas used for consensus.}
\end{figure}


\begin{figure}[t]
    \centering
    \begin{subfigure}[t]{0.6\columnwidth}
	\centering
	\includegraphics[width=\textwidth]{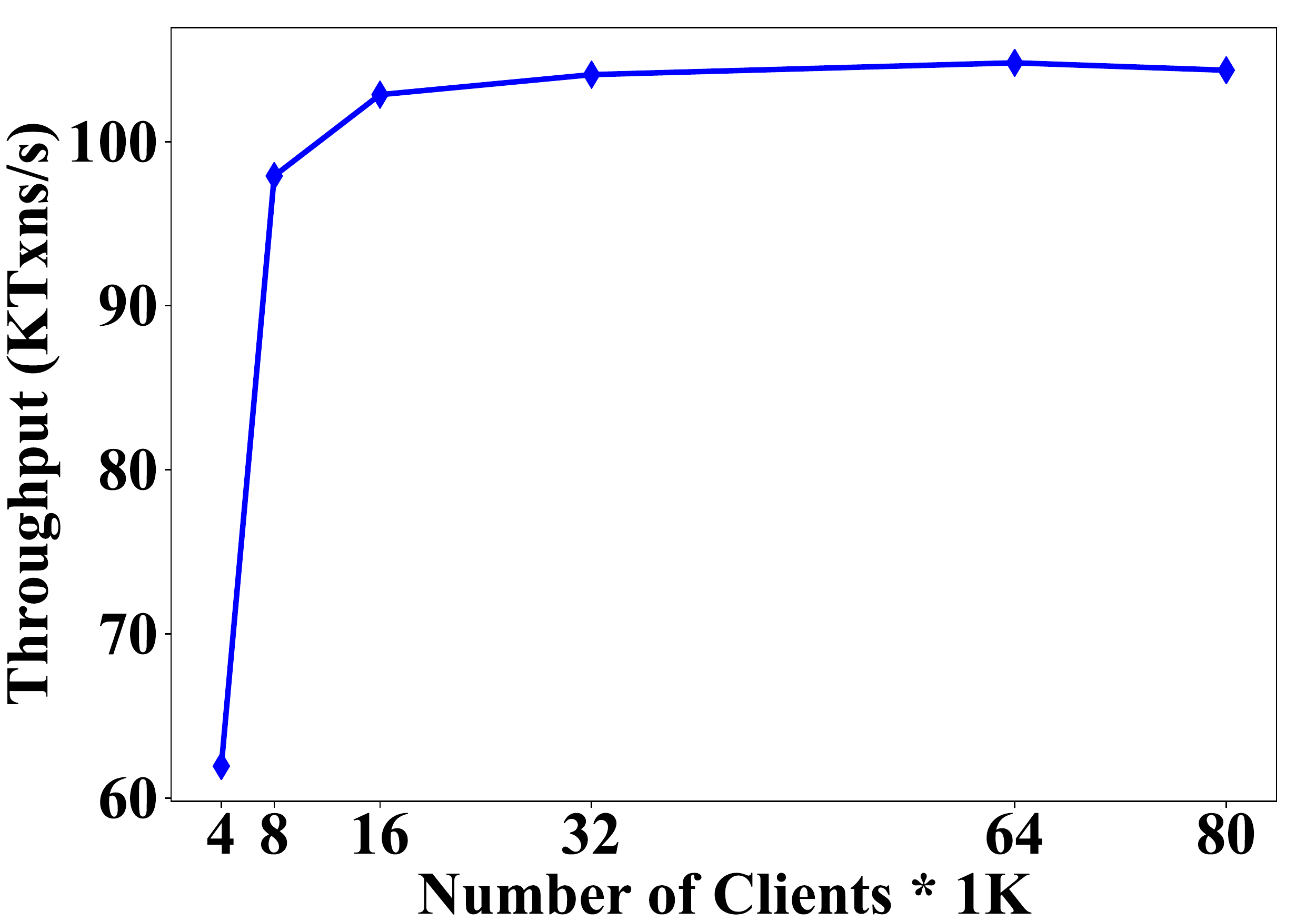}
	\caption{\small System throughput.}
	\label{fig:clients-tput}
    \end{subfigure}
    \begin{subfigure}[t]{0.6\columnwidth}
	\centering
	\includegraphics[width=\textwidth]{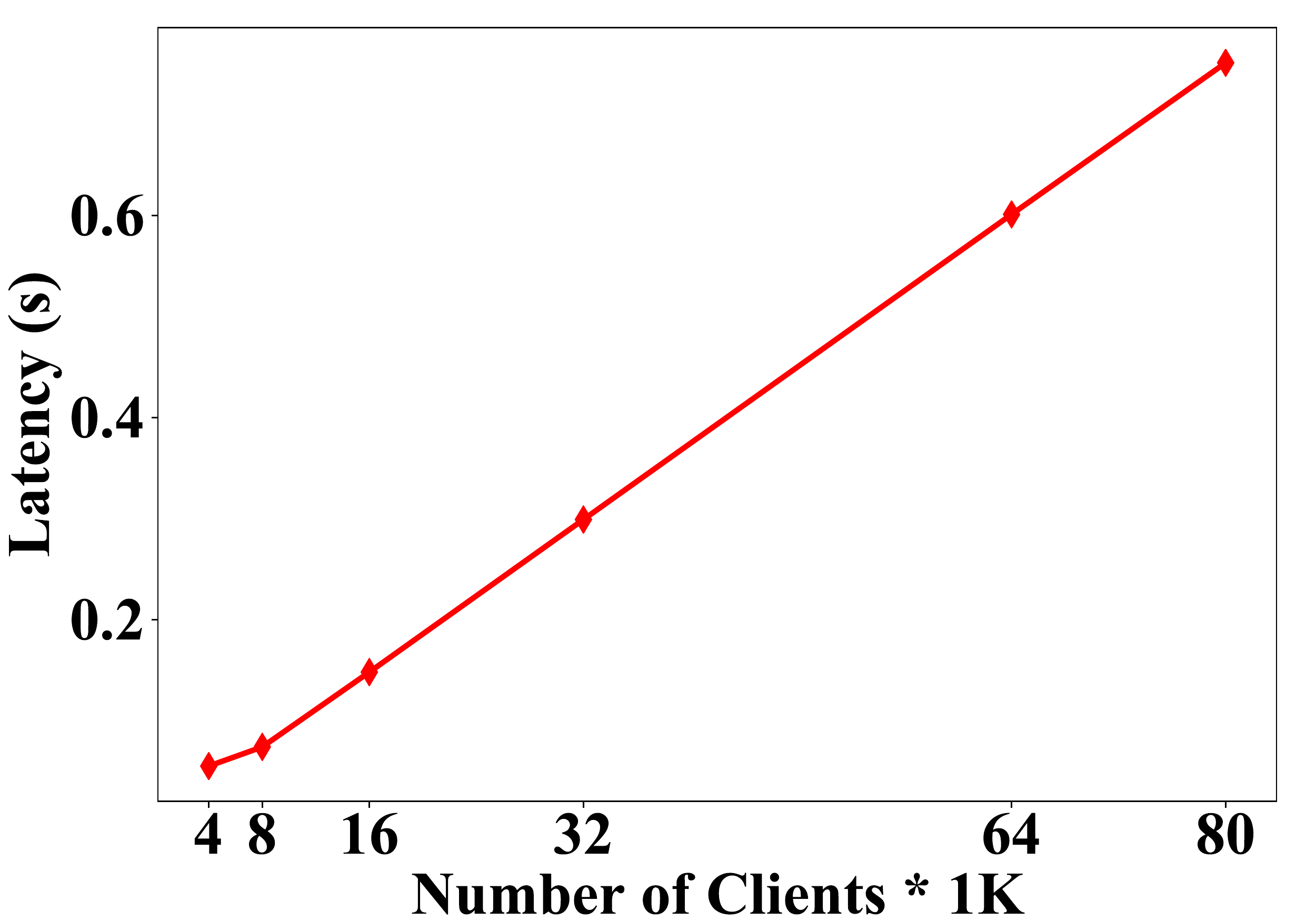}
	\caption{\small Latency.}
	\label{fig:clients-lat}
    \end{subfigure}
      \caption{\small System throughput and latency on varying the number of clients.
Here, $16$ replicas participate in consensus.}
\end{figure}

\subsection{Effect of Memory Storage}
We now try to answer question~\ref{q:mem} by studying the trade-off of having in-memory storage 
versus off-memory storage in a \pbc{}. 
For testing off-memory storage, we integrate SQLite~\cite{sqllite} with our \expodb{} architecture. 
We use SQLite to store and access the transactional records. 
As SQLite is external to our \expodb{} fabric, so we developed API calls to read and write its tables.
Note that until now, for all the experiments, we assumed in-memory storage, that is, records 
are written and accessed in an in-memory key-value data-structure.

In Figures~\ref{fig:mem-tput} and~\ref{fig:mem-lat}, we illustrate the impact on system throughput 
and latency in the two cases.
For the in-memory storage, we require the execute-thread to read/write the key-value data-structure. 
For SQLite, execute-thread initiates an API call and waits for the results. 
It is evident from these plots that access to off-memory storage (SQLite) is quite expensive. 
Further, as execute-thread is busy-waiting for a reply, it performs no useful task.
To {\bf \em summarize:}, choosing SQLite over in-memory storage reduces throughput by $94\%$ and 
increase latency by $24\times$.

\subsection{Effect of Clients}
We now study the impact of clients on a \pbc{} system, and 
as a result, work towards answering question~\ref{q:clients}.
We observe the changes in throughput and latency on increasing the 
number of clients sending requests to a \pbc{} from $4$K to $80$K.

Through Figure~\ref{fig:clients-tput}, we conclude that on increasing the 
number of clients, the throughput for the system increases to some extent (up to $32$K), 
and then it becomes constant. 
This is a result of all the threads processing at their maximum capacities, 
that is, the system is unable to handle any more client requests.  
As the number of clients increases, an increased set of requests have to wait in the queue 
before they can be processed.
This wait can even cause a slight dip in throughput (on moving from $64$K to $80$K clients).
This delay in processing causes a linear increase in the latency 
incurred by the clients (as shown in Figure~\ref{fig:clients-lat}).
To {\bf \em summarize:} we observe that an increase in the number of clients from $16$K to $80$K 
helps the system to gain an additional $1.44\%$ throughput but incurs $5\times$ more latency.


\begin{figure}[t]
    \centering
    \begin{subfigure}[t]{0.58\columnwidth}
	\centering
	\includegraphics[width=\textwidth]{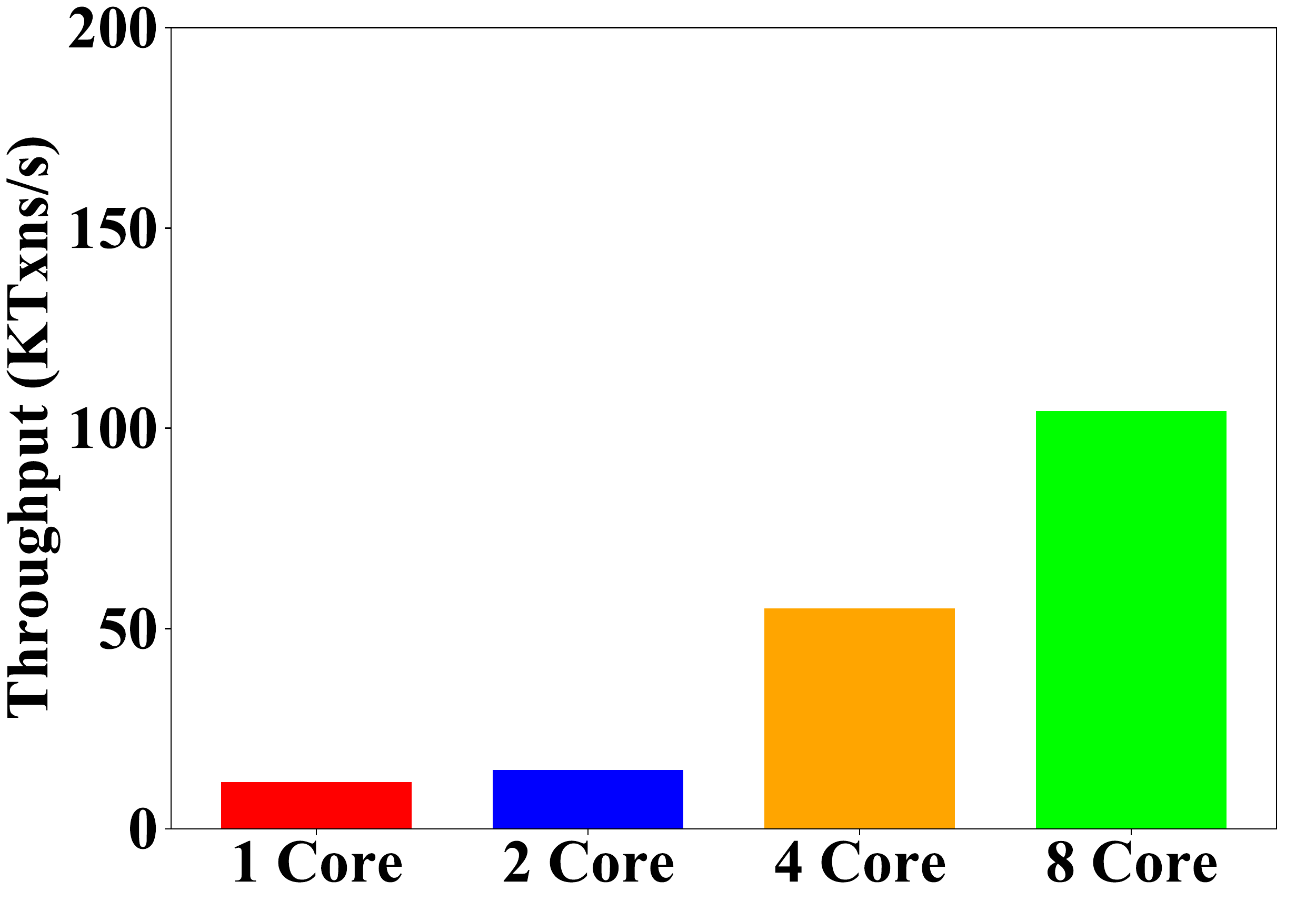}
	\caption{\small System throughput.}
	\label{fig:cores-tput}
    \end{subfigure}
    \begin{subfigure}[t]{0.58\columnwidth}
	\centering
	\includegraphics[width=\textwidth]{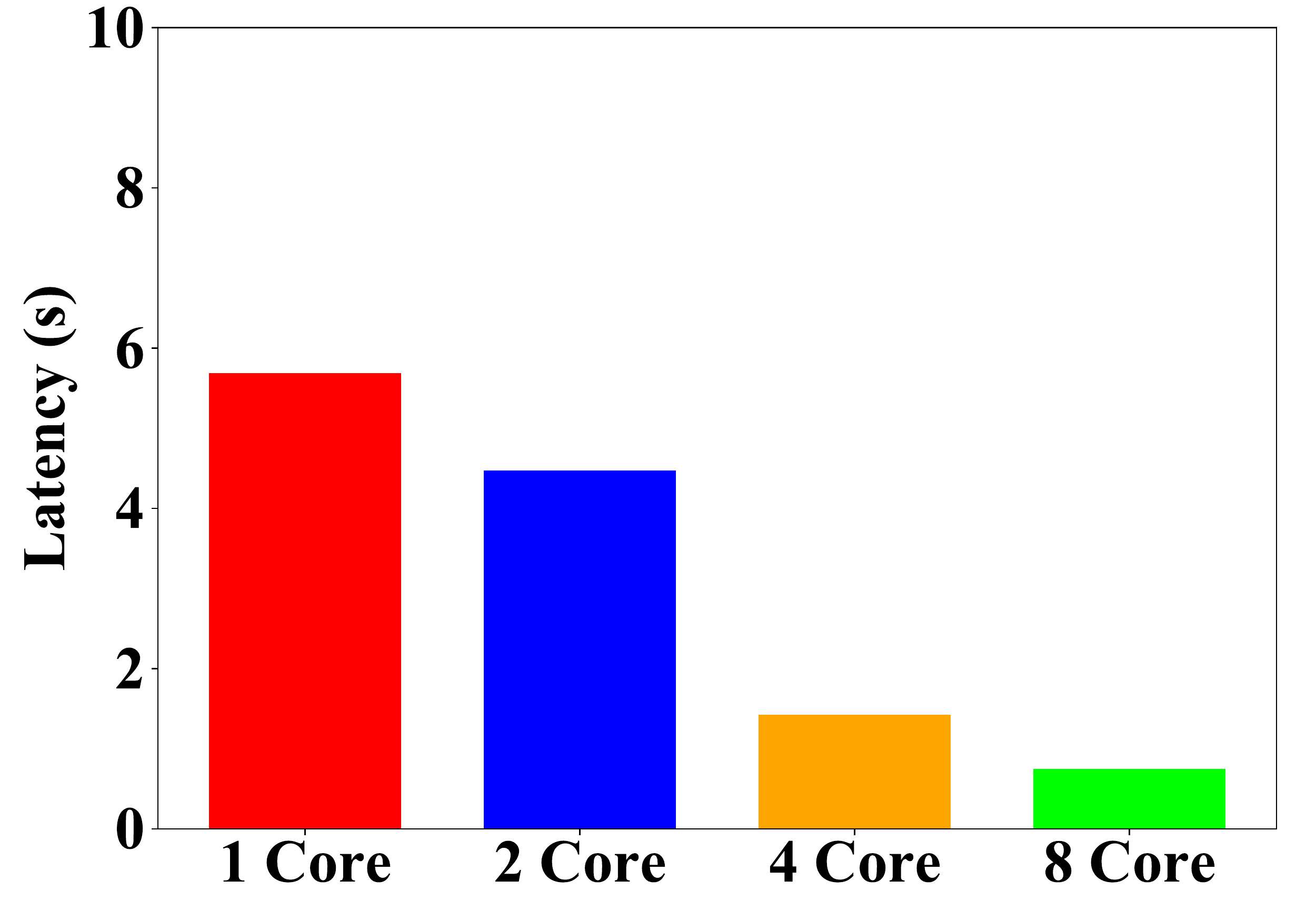}
	\caption{\small Latency.}
	\label{fig:cores-lat}
    \end{subfigure}
      \caption{System throughput and latency on varying the number of hardware cores.
Here, $16$ replicas participate in consensus.}
\end{figure}

\subsection{Effect of Hardware Cores}
We now answer question~\ref{q:cores} by analyzing the effects of 
a deployed hardware on a \pbc{} application.
In specific, we want to deploy our replicas on different Google Cloud machines having 
$1$, $2$, $4$ and $8$ cores.
We use Figures~\ref{fig:cores-tput} and~\ref{fig:cores-lat} to illustrate the throughput and 
latency attained by our \expodb{} system on different machines. 
For all these experiments, we require $16$ replicas to participate in the consensus. 
These figures affirm our claim that if replicas run on a machine with fewer cores, 
then the overall system throughput will be reduced (and higher latency will be incurred). 
As our architecture (refer to Figure~\ref{fig:prim-pipe}) requires 
several threads, so on a machine with fewer cores our threads face resource contention.
Hence, \expodb{} attains maximum throughput on the $8$-core machines.
To {\bf \em summarize:} deploying \expodb{} replicas on an $8$-core machine, in comparison 
to the $1$-core machines, leads to an $8.92\times$ increase in throughput.

\subsection{Effect of Replica Failures}
We now try to answer question~\ref{q:fail} by analyzing whether 
a fast \bft{} consensus protocol can withstand replica failures.
This experiment also illustrates the impact of failures on a \pbc{}.
In specific, we perform  a head-on comparison of \zyzzyva{} against \pbft{}, while allowing 
some backup replicas to fail. 

\begin{figure}[t]
    \centering
    \begin{subfigure}[t]{0.55\columnwidth}
	\centering
	\includegraphics[width=\textwidth]{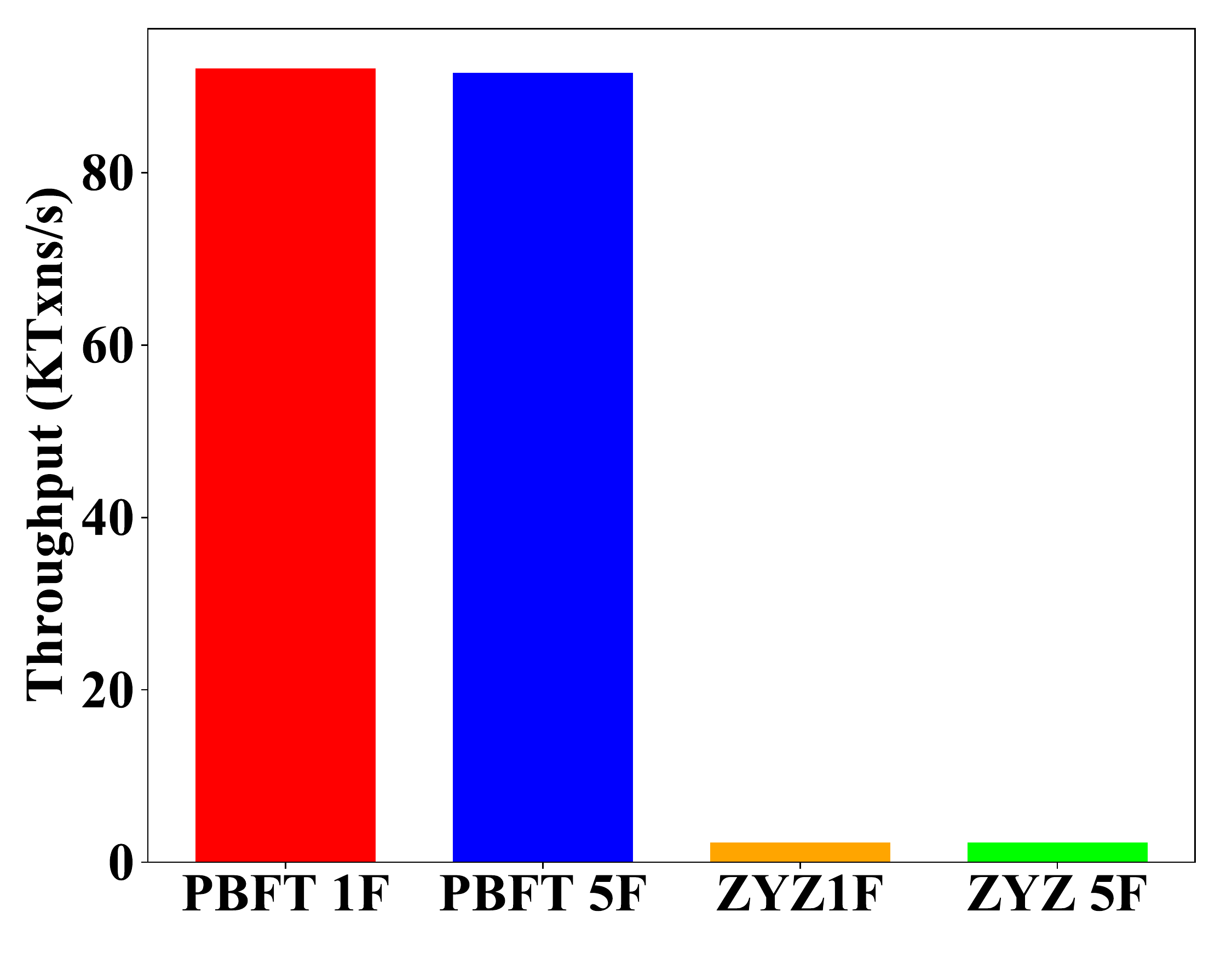}
	\caption{\small System throughput.}
	\label{fig:fail-tput}
    \end{subfigure}
    \begin{subfigure}[t]{0.55\columnwidth}
	\centering
	\includegraphics[width=\textwidth]{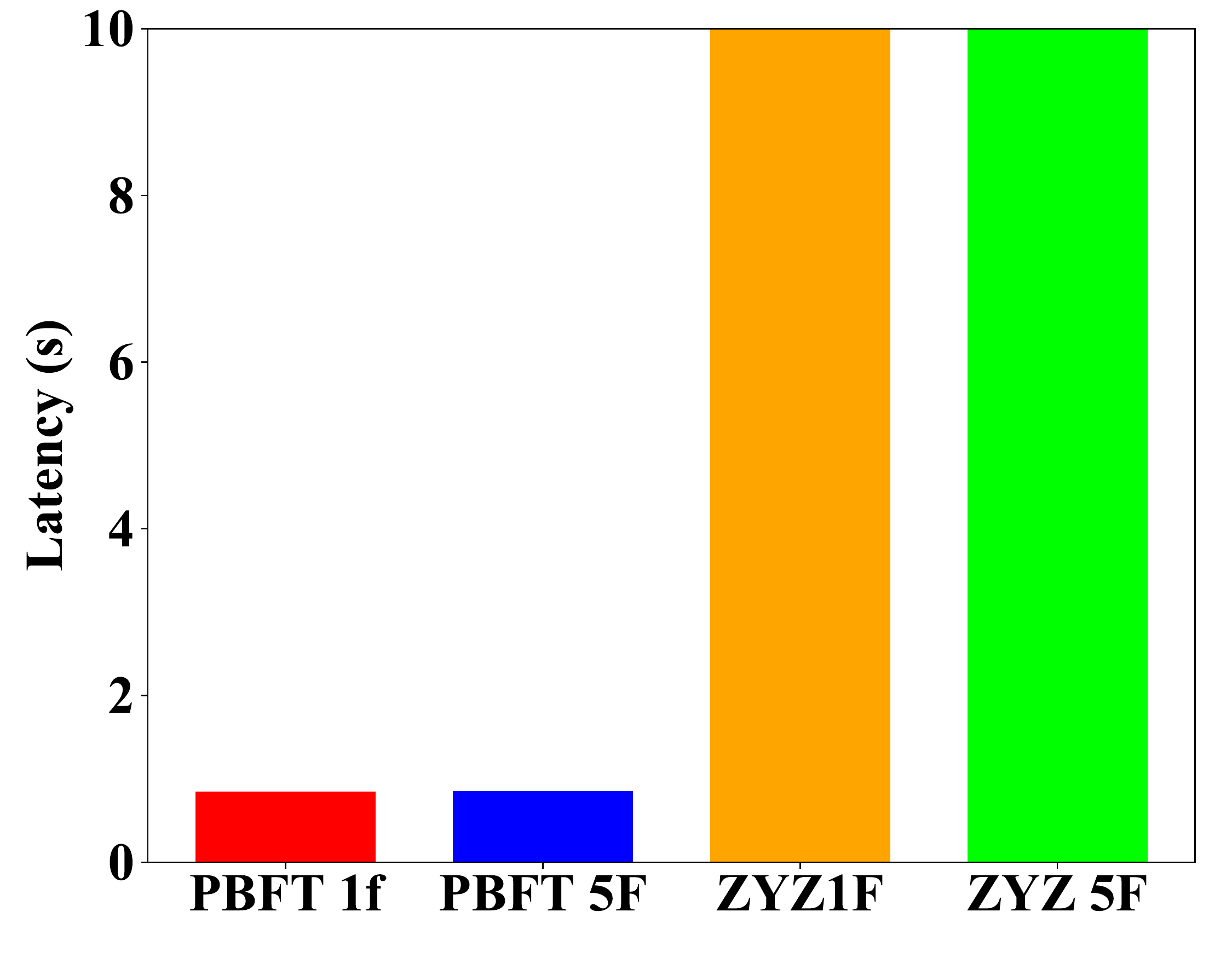}
	\caption{\small Latency.}
	\label{fig:fail-lat}
    \end{subfigure}
      \caption{System throughput and latency on failing non-primary replicas.
Here, $16$ replicas participate in consensus.}
\end{figure}

In Figures~\ref{fig:fail-tput} and~\ref{fig:fail-lat}, we illustrate the impact of 
failure of {\em one} replica and {\em five} replicas on the two protocols.
For this experiment we require at most $16$ replicas to participate in consensus. 
Note that for $n=16$, the maximum number of failures a \bft{} system can handle 
are $f=5$.
Hence, we evaluate both the protocols under minimum and maximum simultaneous failures.

On increasing the number of failures from one to five, 
there is a small dip in the throughput for both the protocols.
This dip is not visible due to the high scaling of the graph. 
For \pbft{}, in comparison to the failure-free case, there is not a significant 
decrease in throughput as none of its phases require more than $2f+1$ messages. 

In case of \zyzzyva{}, the system faces a pronounced reduction in its throughput with 
just one failure. 
The key issue with \zyzzyva{} is that its clients need responses from all the replicas. 
So even one failure makes a client {\em wait} until it {\em timeouts}. 
This wait causes a significant reduction in its throughput.
Note that finding an optimal amount of time a client should wait is a hard problem~\cite{aadvark,upright}.
Hence, we approximate this by requiring clients to wait for only a small time.

Protocols like \zyzzyva{} advocate for a twin path model~\cite{zyzzyva,sbft}.
In these protocols, each replica achieves consensus by following a fast path until the system faces a failure.
Once a failure happens, these protocols decide to switch to a slower path.
Such a design heavily relies on the value of timeout.
If the timeout is large, then these protocols face a large reduction in throughput. 
For example, in \zyzzyva{}, a larger timeout implies clients have to wait for a larger amount of time before 
initiating the next phase.
If the network is dynamic, then the value of timeout can continuously change. 
Thus, finding the optimal value of timeout is hard.
Another way to boost the throughput of these protocols is to assume there are a sufficient number of clients 
that can help offset the effects of timeout.


\section{Observation}
\label{s:observe}
Based on the results presented in the previous section, we 
make two high-level conclusions:
\begin{itemize}
\item A slow classical \bft{} protocol running on a well-crafted implementation (like \expodb{}),
can outperform a fast \bft{} protocol implemented on a protocol-centric design. 

\item No single parameter can alone substantially improve the throughput (or reduce latency) 
of the underlying \pbc{}.
The key reason our \expodb{} framework can attain high throughputs and incurs 
low latency is that it attempts at optimally utilizing several parameters.
\end{itemize}

\textbf{Threading and Pipelining.}
In Section~\ref{s:back}, we discussed several works that 
either present new protocols to improve the
performance of a \pbc{} or illustrate novel use-cases for blockchain.
These works rarely focus on the implementation of a replica itself
and can significantly gain throughput by adopting an architecture similar 
to our \expodb{}.
Further, caution needs to be taken while introducing parallelism as unnecessary threads 
can cause resource contention or deadlocks 
(e.g., multiple execution-threads can cause data-conflicts).

\textbf{Batching and Multiple Operations.} 
Several works suggest batching client requests, while others have vetoed against such a choice.
Our results show that the optimal use of batching can help to reduce the cost of consensus 
by merging multiple consensuses into one. 
However, over-batching does introduce a communication trade-off. 
Hence, each \pbc{} application should determine the optimal set of client requests to batch. 
Clients can also employ multi-operation transactions. 
In practice, such a transaction includes at most ten operations. 
Hence, employing operations per second as a metric to measure throughput may be a good idea. 

\textbf{Message Size and Payload.}
Depending on the application targeted by a \pbc{}, the clients can send requests 
that have a large size. 
For example, a client can require the execution of a specific code. 
If multiple large requests are batched together, then the network may consume resources 
in splitting a message into packets, transmitting these packets, and aggregating these packets at the destination. 
Hence, depending on the application, batching just ten large requests may allow the system to return high throughput.

\textbf{Cryptographic Signatures.}
Although the use of cryptographic signatures bottlenecks the system throughput, 
their use is essential for safety. 
We observe that a combination of MACs and DSs can help guarantee both safety and high throughput.
For instance, digital signatures are only necessary for messages that need to be {\em forwarded}. 
Hence, in a \pbc{}, only clients need to digitally sign their requests. 
For communication among the replicas, MACs suffice, as in most of the \bft{} protocols, 
no replica forwards messages of any other replica. 
Hence, the property of non-repudiation is implicitly satisfied.

\textbf{Chain Storage.}
\pbc{} applications need to store client records and other metadata. 
We observed that the use of in-memory data-structures is better than off-memory storage, such as SQLite. 
The key reason a \pbc{} system can avoid frequent access to off-memory storage is 
that at all times, at most $f$ replicas can fail. 
Hence, if persistent storage is required, then it can be performed asynchronously or delayed until periods of low contention.

\textbf{Replica Failures.}
We know that failures are common. 
Either a replica may fail, or messages may get lost. 
A \pbc{} system needs to be ready to face these situations. 
Hence, the system design must not rely on a \bft{} protocol 
that works well in non-failure cases but attains low throughput under simple failures. 
We observed that designs employing protocols like \zyzzyva{} 
can have negligible throughput with just one failure.
Further, some protocols suggest the use of two modes, fast path and slow path~\cite{sbft}.
Although such protocols attain high throughputs in the fast path, they switch to the slow path on failures. 
Note that this switch happens when some replica or client timeouts. 
Determining the optimal value for timeouts is hard~\cite{aadvark,upright}. 
Thus, twin path protocols may not be suitable if the network is dynamic.

\section{Conclusions}
In this paper, we present a high-throughput yielding permissioned blockchain 
framework, \expodb{}.
By dissecting \expodb{}, we analyze several factors that affect the 
performance of a permissioned blockchain system. 
This allows us to raise a simple question: 
{\em can a well-crafted system based on a classical \bft{} protocol outperform a modern protocol?}
We show that the extensively parallel and pipelined design of our \expodb{} fabric does allow even \pbft{} 
to gain high throughputs (up to $175$K) and outperform common implementations of \zyzzyva{}.
Further, we perform a rigorous evaluation of \expodb{} and illustrate the impact of different 
factors such as cryptography, chain management, monolithic design, and so on.
We envision the practices adopted in \expodb{} to be included in designing and testing newer 
\bft{} protocols and permissioned blockchain applications.

{\bf Acknowledgments.}
We would like to thank DOE Award No. DE-SC0020455 for funding our startup MokaBlox LLC.

\bibliographystyle{IEEEtran}
\bibliography{evalbib}

\end{document}